\documentclass[letterpaper,twocolumn,10pt]{article}
\usepackage{usenix}

\usepackage{tikz}
\usepackage{amsmath}

\usepackage{filecontents}

\usepackage{xspace}

\usepackage{url}
\usepackage{subcaption}
\usepackage{color, colortbl}
\usepackage{siunitx}
\usepackage{breakurl}
\usepackage{enumitem}

\usepackage{algorithm}
\usepackage{algorithmic}
\usepackage{mathtools}

\usepackage{booktabs}
\usepackage{xcolor}
\usepackage{tcolorbox}
\definecolor{mylighterblue}{RGB}{235,245,255}
\definecolor{myorange}{RGB}{255,225,200}
\definecolor{promptblue}{HTML}{6F9BD1}

\usepackage{hyperref}
\usepackage[capitalize,noabbrev]{cleveref}

\hypersetup{
    colorlinks=true,  
    linkcolor=blue,   
    citecolor=blue,  
    urlcolor=blue     
}

\usepackage{fontawesome5}

\usepackage{amsthm}
\newtheoremstyle{definition}%
  {}{}
  {}{} 
  {\bfseries}{.}
  { }{\thmname{#1}\thmnumber{ #2}\thmnote{ (#3)}}
\theoremstyle{definition}

\newcounter{findingcounter}
\newenvironment{finding}{
\refstepcounter{findingcounter}
\begin{tcolorbox}[
colback=mylighterblue,
colframe=white,
boxrule=0pt,
arc=0mm,
left=1mm,
right=1mm,
top=1mm,
bottom=1mm
]
\textbf{Finding~\thefindingcounter:}~%
}{
\end{tcolorbox}
}

\newcounter{patterncounter}
\newenvironment{pattern}{
\refstepcounter{patterncounter}
\begin{tcolorbox}[
colback=myorange,    
colframe=white,
boxrule=0pt,
arc=0mm,
left=1mm,
right=1mm,
top=1mm,
bottom=1mm
]
\textbf{Pattern~\thepatterncounter:}~%
}{
\end{tcolorbox}
}

\usepackage[dvipsnames]{xcolor}
\usepackage{fontenc}
\usepackage{multirow}
\usepackage{listings}
\usepackage{soul}

\newcommand{\para}[1]{\noindent\textbf{#1.}\hspace{.1cm}}

\definecolor{javagreen}{rgb}{0.25,0.5,0.35}

\usepackage{wasysym}
\newcommand{\fullcirc}{\CIRCLE}
\newcommand{\halfcirc}{\LEFTcircle}
\newcommand{\emptycirc}{\Circle}

\usepackage{pifont}

\begin{document}

\newcommand{\benchmark}{\textsc{PatchBench}\xspace}
\newcommand{\benchmarknum}{213\xspace}
\newcommand{\secb}{\textsc{SEC-bench}\xspace}
\newcommand{\cwe}{16\xspace}

\newcommand{\codex}{Codex\xspace}
\newcommand{\aider}{Aider\xspace}
\newcommand{\openhands}{OpenHands\xspace}
\newcommand{\sweagent}{SWEAgent\xspace}
\newcommand{\claudecode}{Claude Code\xspace}

\newcommand{\atlanta}{Atlantis\xspace}
\newcommand{\tob}{Buttercup\xspace}
\newcommand{\theori}{RoboDuck\xspace}

\newcommand{\gptFive}{GPT-5\xspace}
\newcommand{\gptFiveSix}{GPT-5.6 Sol\xspace}
\newcommand{\claudeOpusFourEight}{Claude Opus 4.8\xspace}
\newcommand{\claudeSonnetFourFive}{Claude Sonnet 4.5\xspace}
\newcommand{\geminiThreeFive}{Gemini 3.5 Flash\xspace}
\newcommand{\geminiThreeOne}{Gemini 3.1 Pro\xspace}

\newcommand{\bcircle}[1]{%
   \raisebox{-0.5ex}{\includegraphics[height=2.5ex]{fig/circle#1.pdf}}%
}

\date{}

\title{\Large \bf \benchmark: Evaluating AI Agents for Vulnerability Patching}

\author{
{\rm Chihao Shen}\quad {\rm Jiacheng Li}\quad {\rm Aastha Mahajan}\quad
{\rm Jeffery Siyuan Tian}\quad {\rm Yonghwi Kwon}\quad {\rm Yizheng Chen}\\[4pt]
University of Maryland
}

\maketitle

\begin{abstract}
AI agents have recently demonstrated strong performance in automated vulnerability patching. However, existing evaluations often validate a patch only by testing whether the provided Proof-of-Concept (PoC) input still triggers a crash. This leaves two key threats to validity: agents may reproduce memorized historical developer patches, or they may generate surface-level fixes that only suppress the reported crash.

We study these concerns for C/C++ vulnerability patching. We introduce a patch similarity metric to detect memorized patches. On average, 25\% of the agent patches exhibit substantial similarity to historical developer patches, indicating that patch memorization is a real threat to the validity of vulnerability patching evaluations. Meanwhile, agents also frequently exploit benchmark structures to pass patch validation by patching on the crash stack trace to suppress the crash, rather than localizing and fixing the root cause of the vulnerabilities.

To handle these issues, we propose \benchmark, a new benchmark for evaluating AI agents on realistic vulnerability patching tasks. \benchmark selects vulnerabilities whose ground-truth fixes lie outside the crash stack and uses vulnerability transplant and code mutations to migrate historical vulnerabilities into new repository contexts, reducing the risks of surface-level fixes and patch memorization. We develop new patch validation methods that thoroughly evaluate both security and semantic correctness of agent patches. Across 11 state-of-the-art agents, including the top three AIxCC agents, the original PoC-only validation inflates the patching task solve rate of agents by 1.83$\times$ on average. Our results reveal key limitations of current patching agents and point to future research directions for more reliable vulnerability repair.
\end{abstract}

\section{Introduction}

Patching software vulnerabilities is a crucial, yet expensive and time-consuming security task. Unpatched software vulnerabilities in systems can cost organizations millions of dollars annually~\cite{ibm-data-breach-cost-2025} when exploited by adversaries.
According to Project Zero~\cite{project-zero-metrics-2021}, security vendors took 52 days on average to fix vulnerabilities in 2021.
Recent advances in AI agents have shown a promising path toward automated software vulnerability patching, demonstrated by the AI industry~\cite{openai-aardvark,claude-mythos} and academic events such as the AI Cyber Challenge (AIxCC)~\cite{aixcc-challenge}.
However, rigorous evaluation of AI-assisted patching remains a relevant and challenging problem.

This paper focuses on evaluating AI agents for patching software vulnerabilities, tackling two aspects: memorization of historical security patches and shallow, surface-level patches. First, LLMs underlying the AI agents often memorize their training data~\cite{carlini2022quantifying,carlini2021extracting,yang2024unveiling}, which likely contain bug reports and developer-written patches to known vulnerabilities~\cite{ramos2025large,kong2025demystifying}. 
Hence, evaluation using known vulnerabilities~\cite{lee2026sec} may end up testing memorized knowledge extraction, instead of agents' analysis and patching capabilities. Note that AIxCC has 63 manually written synthetic vulnerabilities (40 in C and 23 in Java)~\cite{zhang:sok} to mitigate the memorization issue. 
Unfortunately, such a manual effort is expensive and not scalable for evaluating AI agents that patch vulnerabilities across diverse real-world projects.

Second, prior automatic patch validation methods fail to assess the semantic correctness of agent patches. Prior work often relies on testing against a few inputs, such as Proof-of-Concept (PoC) inputs and project-level functional test cases.
Hence, they often fail to validate incomplete, surface-level fixes, frequently generated by agents that take shortcuts and guess at solutions~\cite{kapoor2026holistic}. Worse, they also fail to identify patches altering critical functionalities (e.g., patching by removing functionality containing vulnerabilities), leading to inflated performance. A study~\cite{zhang:sok} reveals that Team Atlanta of AIxCC and Claude Code can pass all automatic patch validation, but 16\% to 38\% of patches they generated are still semantically incorrect.

In this paper, we first focus on measuring the impact of LLM memorization on patching. Specifically, we study the extent to which security patches are memorized by LLMs and how this affects AI agents. Note that while prior works have explored LLMs memorizing bug benchmarks~\cite{ramos2025large,kong2025demystifying}, their methods focus on function-level program repair tasks, which are not directly applicable to general security patching at the repository level. To this end, we develop a new similarity metric that aims to detect memorized patches generated by LLMs. We then conduct a study under two distinct settings using the state-of-the-art vulnerability patching benchmark \secb~\cite{lee2026sec} containing 300 patching tasks in C/C++: (1) local-context LLM-based patching and (2) repository-level agent-based patching. Surprisingly, we find that agents Codex, Claude Code and OpenHands produce a higher proportion of likely memorized patches, compared to the underlying LLM alone. Specifically, we find that an average of 25\% of agent patches are extremely similar to, if not the same as, the historical ground-truth patches written by developers. We hypothesize that agent scaffolds provide better code context, which could enable the underlying LLM to produce more memorized content.

Moreover, we observe that AI agents' patches frequently appear in functions on the crash stack trace, regardless of the vulnerable code's location. Overall, 81\% of \codex + \gptFiveSix's patches modify functions on the crash stack trace. Even when the root cause of the vulnerability is \emph{not located} in the functions in the crash stack trace as indicated by the historical developer patch, 64\% of such agent patches are still found in functions on the trace. These patches may only suppress the crash of a single PoC but do not fix the root cause of the vulnerability. Therefore, PoC input-based benchmarks (e.g., in \secb) cannot precisely reflect the patches' validity.

To this end, we propose a rigorous vulnerability patching benchmark \benchmark{}, which comprehensively evaluates the capabilities of AI agents in generating quality security patches, including \benchmarknum patching tasks across 16 CWEs and 32 real-world projects. Specifically, to tackle AI agents relying on the crash stack trace for localizing vulnerabilities, we focus on vulnerabilities whose ground-truth patches do not appear in functions of the crash stack trace. They help differentiate agents that take shortcuts from those that conduct localization. In addition, to measure agents' capability beyond patch memorization, we transplant historical vulnerabilities to newer versions of their repositories and mutate the code at the patch sites, so that agents face unseen context and cannot reproduce a historical patch verbatim. Finally, we manually audit the reference patch of every task. We find that commits labeled as security fixes do not always eliminate the root cause, and many carry code changes unrelated to the vulnerability~\cite{jiang2022Evocatio, he2023SVEN}. We therefore discard tasks of the former kind and strip the unrelated changes from the latter, so that the reference patch of each task is a manually curated fix for the root cause rather than the original developer patch.

Furthermore, we validate patches in both security and semantic aspects.
Our \emph{security validation} checks whether a patch eliminates the target vulnerability on fuzz-generated crashing inputs in the vulnerable repository, while our \emph{semantic validation} checks whether a patch preserves expected behavior on benign inputs and project unit tests against the reference-patched repository. A patch is considered semantically valid if and only if it satisfies all three of the following conditions: (1) The agent-patched repository does not trigger sanitizer errors on benign inputs. (2) The program-level output state on benign inputs match that of the reference-patched repository. (3) It passes all working unit tests from the reference-patched repository.
Compared to our work, the most comprehensive patch validation techniques from AIxCC teams do not use program-level output state validation for benign inputs, and not all teams check multiple PoCs, which could lead to semantically incorrect patches~\cite{zhang:sok}.

\begin{figure}[t!]
\centering
\includegraphics[width=\columnwidth]{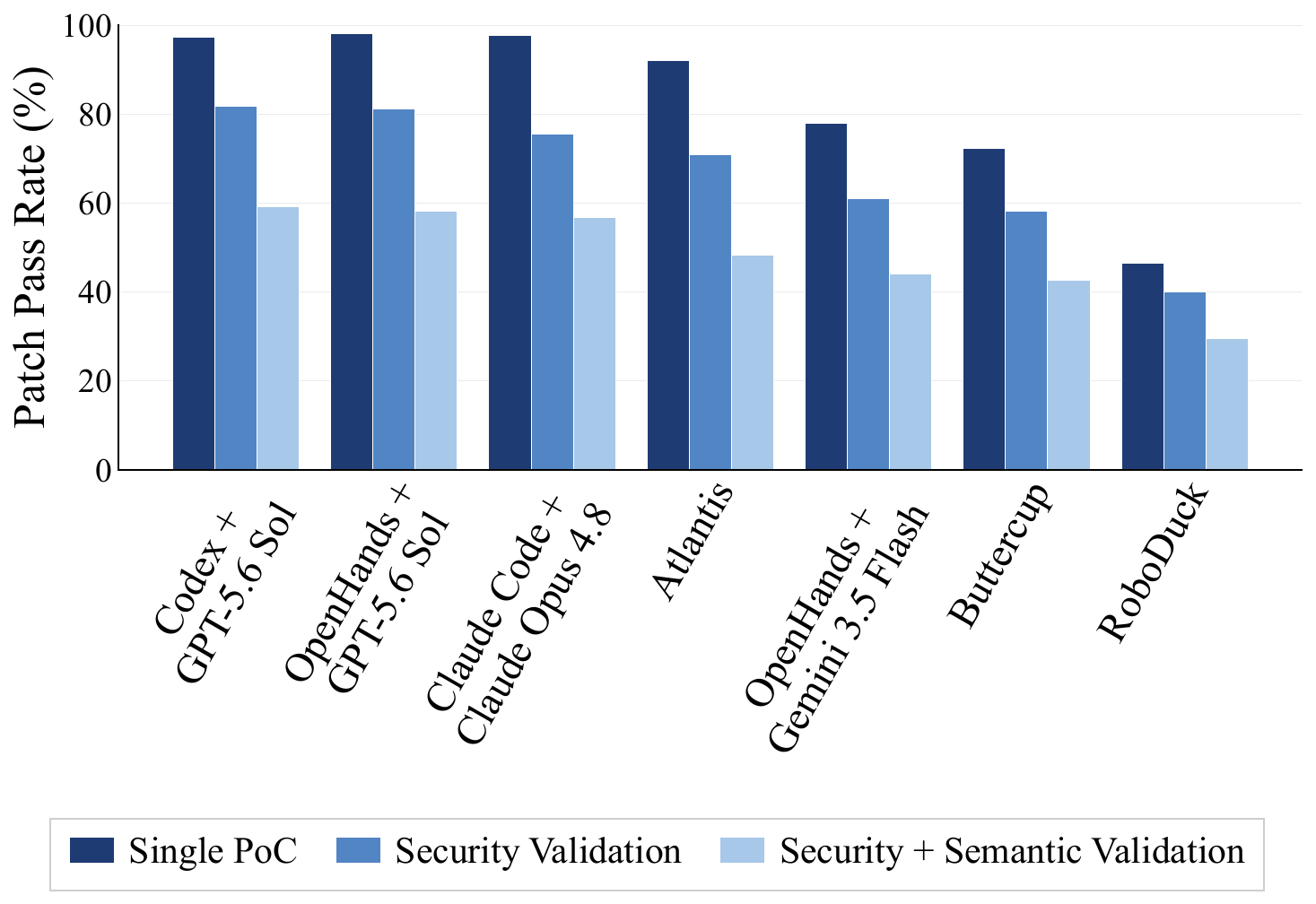}
\caption{The top three agents achieve Single-PoC pass rates above 97\%, while their pass rates drop to 75–82\% after security validation with multiple PoCs, and to roughly half with additional semantic validation, which checks whether patches preserve intended program behavior.
}
\label{fig:intro_result}
\end{figure}

\begin{figure*}[!t]
\centering
\includegraphics[width=.95\linewidth]{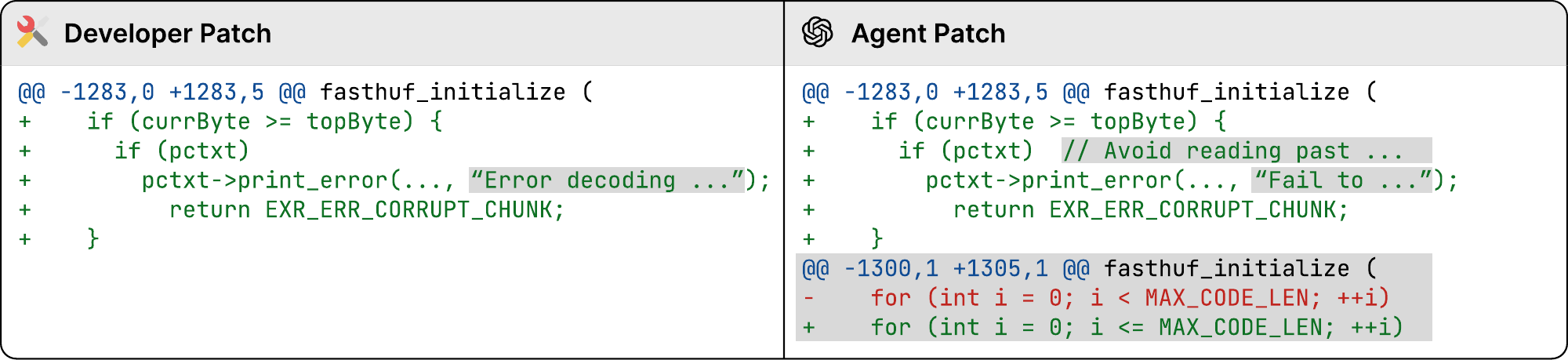}
\caption{
Comparison between the developer patch and the agent patch for OSS-Fuzz bug 42517450 in \texttt{openexr}. \codex{} generates a patch that matches the developer patch at the same program location and introduces the same guard condition, despite superficial textual differences in the comment, error message, and an additional diff hunk.}
\label{fig:mem_example}
\end{figure*}

We evaluate \benchmark on 11 state-of-the-art patching agents, including the top three AIxCC agents~\cite{aixcc-challenge} and 8 general-purpose agents spanning commercial scaffolds, such as Codex \cite{openai_codex} and Claude Code \cite{Anthropic2024}, and an open-source scaffold, OpenHands \cite{wang2025openhands}, with different large language models. As shown in \autoref{fig:intro_result}, PoC-only validation inflates solve rates by 1.83× on average, and even the strongest agents are affected: the top three pass over 97\% of the original PoCs but solve roughly half of tasks under our Security + Semantic Validation.
We observe that the current patching agents frequently fail to preserve functional correctness on valid inputs.
\benchmark contains 67 patching tasks that are not solved by any of the 11 agents, regardless of the patching budget allocated to LLMs (performance plateaus well before a 5$\times$ budget cap; see \autoref{sec:results}). We make the following contributions:

\begin{itemize}[itemsep=0pt, topsep=0pt]
\item We study security patch memorization of AI agents and large language models, using a new patch similarity metric. We find that memorization is a critical challenge for the evaluation of patching agents.
\item We propose \benchmark, a new benchmark to evaluate AI agents for patching C/C++ vulnerabilities, that mitigates the vulnerability memorization issue.
\item We propose patch validation techniques,  comprehensively covering both security and semantic validation.
\item We thoroughly evaluate 11 state-of-the-art patching agents including commercial agents and top three AIxCC agents. Our results reveal bottlenecks and provide insights for developing stronger patching agents.
\item We release our code and benchmark at \url{https://github.com/ai-sec-lab/PatchBench}.
\end{itemize}

\section{Motivation}

We use a recent benchmark, \secb~\cite{lee2026sec}, to study the current patching ability of AI agents. \secb is a repository-level benchmark for evaluating AI agents on realistic C/C++ vulnerability repair. It contains 300 tasks derived from OSS-Fuzz~\cite{ossfuzz} and CVE~\cite{nist_nvd}, each providing a reproducible vulnerable codebase, sanitizer report, triggering PoC, and validation commands. These artifacts allow agents to modify and compile the program and test their patches.

Although \secb is designed to require vulnerability localization, contextual reasoning, and dependency-preserving repair, recent agents perform surprisingly well under its original setup and validation procedure. With a maximum budget of \$5 per task, our evaluation shows that \codex with \gptFiveSix passes 97.3\% of the tasks. Rather than necessarily demonstrating robust repair capabilities, this result raises two concerns. First, because the tasks are based on public vulnerabilities, agents may reproduce memorized developer patches instead of deriving fixes through reasoning. Second, passing a single provided PoC does not guarantee a robust patch: a patch may merely suppress the observed crash or inadvertently break benign functionality.

\subsection{Data Contamination Issue}

\autoref{fig:mem_example} gives an example developer patch from \secb and the corresponding agent patch generated by \codex. The agent patch is quite similar to the developer patch, but not textually identical, which suggests a potential data contamination issue. The developer patch adds a boundary check and prints some error messages. The agent patch reproduces the same guard at the same program location, with the same control-flow structure and return value. However, as highlighted in the grey background, the agent patch adds a new comment, prints slightly different error messages, and also changes a loop bound (in the second patch hunk).

If we directly compare the tokens in the two patches, superficial differences such as a different string literal in the error message and an appended comment make them look quite different, even though they are indeed similar. To address this, we propose a new tokenizer in \autoref{sec:memorization}. 
Moreover, existing similarity metrics are not suitable for code diffs since they do not directly model multiple hunks in two patches. This example illustrates why patch memorization cannot be measured reliably by applying a generic text similarity metric directly to patch hunks, motivating our patch similarity detection method defined in \autoref{sec:memorization}.

Critically, we need to understand how often such memorization behavior occurs in \secb. Since \secb is built from public vulnerabilities, a high pass rate may reflect large language models producing memorized fixes that they have trained on from public repositories. Therefore, we conduct a study to quantify the fraction of agent patches highly similar to historical developer patches, potentially suggesting the memorization effect.

\begin{figure}[!t]
    \centering
    \begin{subfigure}{0.9\linewidth}
        \centering
        \includegraphics[width=\linewidth]{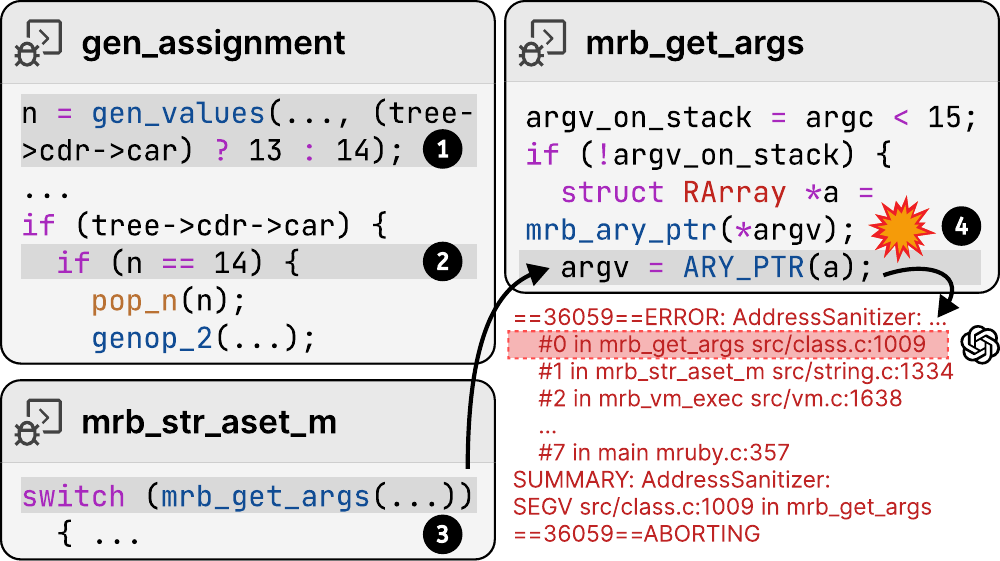}
        \caption{Compiler-side argument packing vulnerability that later triggers a VM-side OOB read.}
        \label{fig:vulnerability}
    \end{subfigure}

    \vspace{0.5em}

    \begin{subfigure}{0.9\linewidth}
        \centering
        \includegraphics[width=\linewidth]{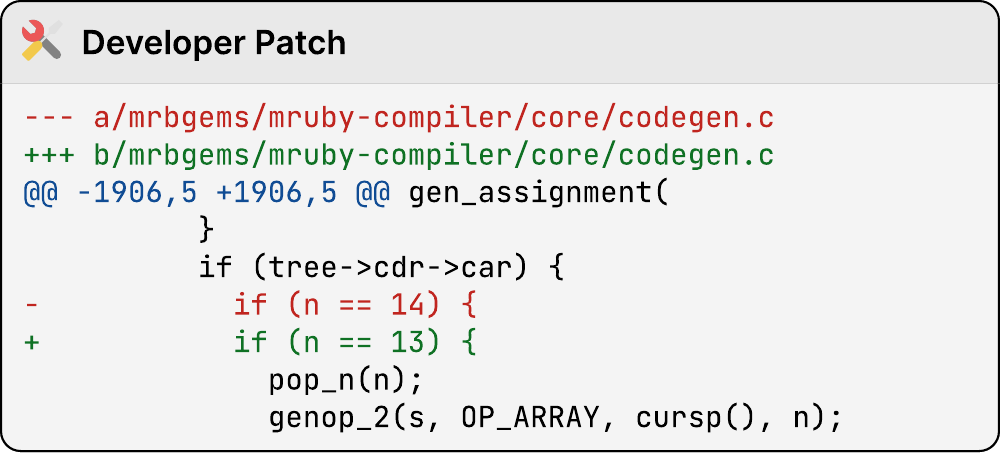}
        \caption{Developer Patch}
        \label{fig:developer_patch}
    \end{subfigure}

    \vspace{0.5em}
    
    \begin{subfigure}{0.9\linewidth}
        \centering
        \includegraphics[width=\linewidth]{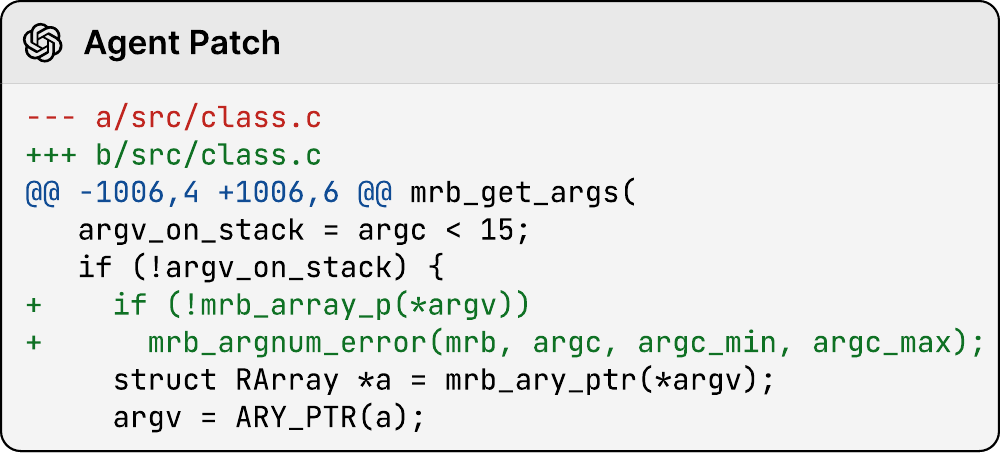}
        \caption{Agent Patch}
        \label{fig:agent_patch}
    \end{subfigure}
    \caption{CVE-2022-1276 in \texttt{mruby}. The developer patch fixes the compiler-side argument packing condition at the correct location, while the agent adds only a surface-level check at the VM crash location.}
    \label{fig:validation}
\end{figure}

\subsection{Patch Validation Issue}
\label{sec:Patch Validation Issue}

\autoref{fig:vulnerability} illustrates a real-world vulnerability CVE-2022-1276, an Out-of-Bounds (OOB) read in \texttt{mruby}, from \secb. This example explains why passing the provided PoC is not sufficient for patch validation. 

\para{Background}
In Ruby, a program is first compiled to bytecode and then executed on its virtual machine, where a vulnerability may involve both the compiler and the virtual machine.
Specifically, the compiler (in \texttt{compiler.c}) generates bytecode from the source code. In particular, for Ruby function calls with 14 or fewer arguments, the arguments are placed directly on the register stack. If there are 15 or more arguments, the compiler needs to pack them into an array and place that array in a single register. Second, the register-based virtual machine (VM) (in \texttt{vm.c}) executes the compiled bytecode. The function \texttt{mrb\_get\_args} from the VM side checks the argument count to decide whether to read the arguments out of an array. In this example, the AddressSanitizer report points to a crash on the VM side, but the root cause of the vulnerability is in the compiler.

\para{Vulnerability Root Cause}
The vulnerability originates from an inconsistency in the compiler's argument packing logic. The arguments consist of positional arguments and keyword arguments. As shown in \autoref{fig:vulnerability}, in the compiler function \texttt{gen\_assignment}, when \texttt{tree->cdr->car} is \texttt{true}, the compiler sets $n=13$ to indicate that argument packing should happen once there are 13 positional arguments (\bcircle{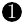}), since the keyword argument and the assigned value occupy two additional argument slots and bring the total argument count to 15. However, the later packing condition check uses an incorrect constant \texttt{14} at \bcircle{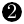}. 
Observe that at \bcircle{1}, \texttt{14} is assigned when \texttt{tree->cdr->car} is \texttt{false}, which conflicts with the predicate guarding \bcircle{2}.

On the VM side, when \texttt{mrb\_str\_aset\_m} processes arguments for the Ruby string slice assignment method using \texttt{mrb\_get\_args} (\bcircle{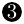}), it sees that the number of arguments is 15, expecting that arguments have been packed. As a result, \texttt{mrb\_get\_args} tries to read the unpacked argument scalar value register as an array pointer, causing the OOB read (\bcircle{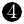}).

\para{Developer Patch}
\autoref{fig:developer_patch} shows the developer patch, which modifies the compiler-side packing condition in \texttt{gen\_assignment}, making sure that argument packing indeed occurs when it should. This is a desirable patch as it eliminates the root cause of the vulnerability \emph{in the compiler}. The malformed state is introduced during bytecode generation, before execution ever reaches the VM argument parser.
However, the sanitizer crash stack can only see the VM's failure (in the argument parsing routine), making it impossible to connect the crash to its root cause.

\para{Agent Patch}
Rather than patching the root cause in the  compiler, the agent adds a surface-level array-type check at the crash point in the VM (\autoref{fig:agent_patch}). This successfully stops the reported crash in the Ruby string slice assignment method, but the logical bug remains, transforming it into a silent error. In other words, the fix remains incomplete and the patched program can still silently generate malformed outputs.

\textit{Incompleteness.} The agent patch guards only one of the several argument-reading routines in the VM. Other routines that parse arguments are not patched, so the vulnerability remains, e.g., similar argument assignments to the Array slicing method still result in an OOB read.

\textit{Malformed Functionality.} More seriously, the agent patch fails to repair the malformed behavior silently generated by non-crashing inputs. In the agent-patched version, the argument packing logic is still wrong. If the first argument of \texttt{argv} is a valid array for string slice assignment, the check \texttt{mrb\_array\_p(*argv)} succeeds, and the guard is bypassed. The code silently proceeds to unpack that array as if it were the entire argument list, and the method simply uses that wrong set of arguments to generate the answer.

This example reflects a broader pattern: 81\% of patches generated by \codex + \gptFiveSix modify a function on the crash stack. Even among the 92 tasks whose developer patches lie outside the crash stack, 64\% of agent patches modify crash-stack functions while passing the PoC. This suggests that agents often bypass vulnerability localization, while single-PoC validation rewards patches that suppress the observed crash without fixing its root cause.

These findings motivate \benchmark and our validation design (\autoref{sec:validation}). \benchmark selects tasks whose developer patches lie outside the crash stack trace, requiring agents to reason about the vulnerability in a broader context. Our patch validation uses related crashing inputs to test whether the sanitizer error is truly eliminated, and benign inputs to check that the patched program preserves benign behavior.

\begin{figure*}[!t]
\centering
\includegraphics[width=\linewidth]{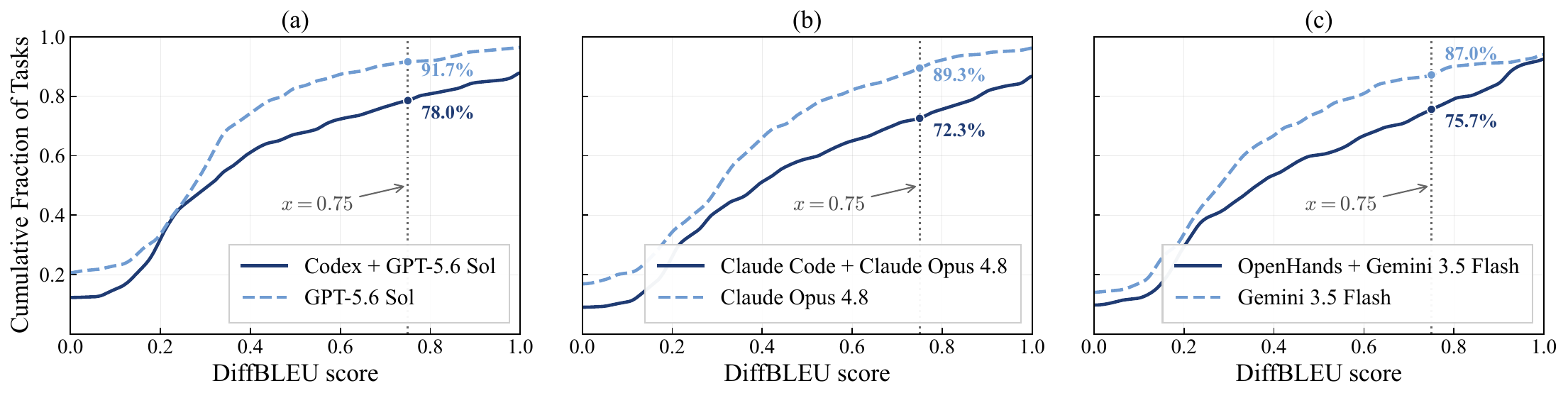}
\caption{
DiffBLEU similarity between agent patches and developer patches on \secb. Agents consistently shift the score distribution toward higher developer patch similarity compared with their corresponding standalone LLMs.
}
\label{fig:cdf}
\end{figure*}

\section{Patch Memorization}
\label{sec:memorization}
Data contamination can influence AI agents in code generation. In this section, we focus on one concrete and measurable way: \textit{whether agents generate patches by reproducing historical developer fixes}. We refer to this behavior as \textit{patch memorization}. We do not  aim to prove other contamination effects, as the training data of many modern language models is unavailable, and data contamination can appear in different forms. Instead, we concentrate on patch-level evidence, where we measure the similarity between generated patches and developer patches.

\subsection{Detection Methods}
\label{subsection:Detection Methods}

There is no universal detection method that determines whether two patches are similar. Prior work commonly uses different methods, including n-gram accuracy, negative log-likelihood for code benchmark leakage detection~\cite{ramos2025large}, and exact match, PPL-based methods for code generation~\cite{yang2024unveiling,kong2025demystifying}. Other general memorization studies use textual-overlap metrics such as ROUGE-based methods~\cite{golchin2024time,yao2024large}. However, these methods suffer from the following two problems.

\para{Code Awareness}
A patch is not plain text, but it combines program structure with edit operations.
Token-level metrics ignore syntactic roles and may fail to capture similarities
in templates or control-flow constructs such as \texttt{if}, \texttt{switch},
and \texttt{while}. They also do not distinguish edit directions, e.g.,
an added line and a removed line may be treated similarly even though they have
opposite meanings. Moreover, such metrics are sensitive to superficial changes
such as formatting, comments, and literals. As a result, semantically similar
patches may receive low scores, while patches with overlapping tokens but
different edits or behaviors may receive high scores.

\para{Context Awareness}
A hunk alone does not fully characterize a patch. The behavior of a changed
statement often depends on its surrounding control and data flow, especially in
C/C++, where branches, loop conditions, variable definitions, and macros can
determine the meaning of the edit. At the same time, comparing the entire
function or file can introduce large amounts of unchanged code that dominate the
similarity score. A good detection method should therefore compare the changed
hunks together with only the relevant local context.

\subsection{DiffBLEU}

We introduce \textbf{DiffBLEU}, a context-aware patch similarity metric built on top of the code-aware design of CodeBLEU~\cite{ren2020codebleumethodautomaticevaluation}. Before describing DiffBLEU, we briefly summarize the CodeBLEU score here. Given the reference code snippet $x_r$ and the candidate code snippet $x_c$, CodeBLEU computes how well $x_c$ matches the ground truth $x_r$:

\begin{equation}
\begin{aligned}
{}&{}\mathrm{CodeBLEU}
= 
\alpha \, \mathrm{BLEU}(x_r,x_c) 
+ \beta \, \mathrm{BLEU}_{\mathrm{w}}(x_r,x_c) \\
&+ \gamma \, \mathrm{Match}_{\mathrm{ast}}(x_r,x_c)
+ \delta \, \mathrm{Match}_{\mathrm{df}}(x_r,x_c)
\end{aligned}
\end{equation}

\noindent where $\mathrm{BLEU}$ is the standard $\mathrm{BLEU}$ score~\cite{papineni2002bleu} that calculates the n-gram-based precision,
$\mathrm{BLEU}_w$ is a weighted variant of $\mathrm{BLEU}$ that assigns higher weights to keywords than to other tokens,
$\mathrm{Match}_{\mathrm{ast}}$ calculates the percentage of candidate AST subtrees that match those in the reference code, and 
$\mathrm{Match}_{\mathrm{df}}$ calculates the fraction of the data-flow edges in the candidate code that match those in the reference code.

DiffBLEU adapts CodeBLEU by comparing both tokens in code patches and the context around the patches, incorporating our diff-aware tokenizer and program slices.
Let $\Delta_r$ be the developer patch, $\Delta_c$ be the agent patch, and let
$C_r$ and $C_c$ be the corresponding code contexts containing the
code diff hunks. We define DiffBLEU as:

\begin{equation}
\begin{aligned}
{}&{}\mathrm{DiffBLEU}
=
\alpha \, \mathrm{BLEU}(\Delta_r,\Delta_c)
+ \beta \, \mathrm{BLEU}_{\mathrm{w}}(\Delta_r,\Delta_c) \\
&+ \gamma \, \mathrm{Match}_{\mathrm{ast}}(C_r,C_c)
+ \delta \, \mathrm{Match}_{\mathrm{df}}(C_r,C_c).
\end{aligned}
\end{equation}

\noindent where $\mathrm{BLEU}$ and $\mathrm{BLEU}_{\mathrm{w}}$ are
computed using a diff-aware tokenizer over code diffs $\Delta_r$ and $\Delta_c$, while
$\mathrm{Match}_{\mathrm{ast}}$ and $\mathrm{Match}_{\mathrm{df}}$ are computed
on the code contexts $C_r$ and $C_c$.
The first two terms calculate the surface-level similarity of the patch hunks. The last two terms capture control-flow and data-flow structure similarity over patch context.

\para{Diff-aware Tokenizer} We construct a diff-aware tokenizer to preserve the edit operation, such that tokens from added lines and removed lines are mapped into separate spaces. Apart from that, the tokenizer also normalizes code-irrelevant noise by anonymizing literals, removing comments,  preserving code-specific tokens, etc.

\para{Code Context for AST}
We compute $\mathrm{Match}_{\mathrm{ast}}$ over control-flow slices in the code contexts that contain the patches.

\para{Code Context for Data Flow}
We compute $\mathrm{Match}_{\mathrm{df}}$ over data-flow slices in the code contexts related to data dependency of variables used in the patches, including variable definitions and dependent statements.

A higher score of DiffBLEU indicates that the agent patch is more similar to the historical patch written by the developer. This does not necessarily mean the patch is correct. We use DiffBLEU to detect memorized patches. We set the hyperparameters to $\alpha=0.3$, $\beta=0.3$, $\gamma=0.2$, and $\delta=0.2$, giving slightly lower weight to the context scores, and a similarity threshold that considers a generated patch with $\mathrm{DiffBLEU} > 0.75$ as a memorized patch. Our manual analysis over a validation set shows that these hyperparameters result in the lowest false omission rate (1.1\%) with no false positives. The detailed validation results are in \autoref{app:mem}.

\subsection{Memorization Study}

\begin{figure*}[t!]
\centering
\includegraphics[width=.95\textwidth]{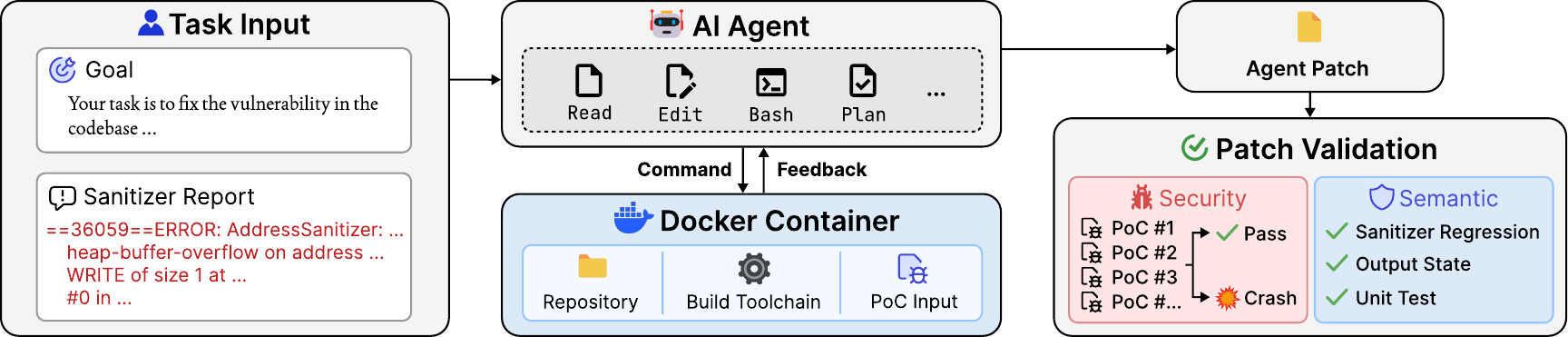}
\caption{Overview of \benchmark. Each task provides the agent with a repository-level repair environment containing the task repository, build toolchain and the PoC input. The agent modifies the source code within a Docker container, after which the generated patch is evaluated with security validation on crashing inputs and semantic validation on benign inputs.}
\label{fig:overview}
\end{figure*}

We use our new detection method to test whether the high \secb pass rate (97.3\%) is accompanied by patch memorization.

\para{Experimental Setups} We consider two settings.

\begin{enumerate}[label=(\arabic*),leftmargin=*, itemsep=0pt, topsep=0pt]
\item \emph{Local-context LLM-based Patching.} Following prior studies~\cite{pearce2023examining, kong2025demystifying,ramos2025large}, for each task, we extract the vulnerable code identified by the developer patch, including the enclosing entity, e.g. function or structure, together with header macros, and ask the model to generate a patch given this context.
\item \emph{Repository-level Agent-based Patching.} We use AI agents to generate the patch. Unlike the local-context setting, the agent has access to the full repository, sanitizer report, PoC, build configuration, and validation commands.
\end{enumerate}

We strictly follow the \secb setup for repair instructions. Each task prompt provides the sanitizer report, the location of the repository, and the same step-by-step repair requirements. In the local-context setting, we evaluate \gptFiveSix, \geminiThreeFive, and \claudeOpusFourEight. Each model generates a patch without repository access or execution feedback. In the repository-level setting, we choose \codex + \gptFiveSix, \openhands + \geminiThreeFive, and \claudecode + \claudeOpusFourEight. All experiments use a maximum budget of \$5 per task and medium reasoning effort.

\para{Memorization Measurement Results} In \autoref{fig:cdf}, each curve represents the cumulative distribution of DiffBLEU scores between generated patches and developer patches on \secb. For all three model families, repository-level agents produce patches that are more similar to developer patches than standalone LLMs do. At the memorization threshold of 0.75, \gptFiveSix has 8.3\% of patches above the threshold, while \codex + \gptFiveSix has 22.0\%. The same trend appears for both \claudeOpusFourEight and \geminiThreeFive, where the fraction above the threshold increases from 10.7\% to 27.7\% under \claudecode, and from 13.0\% to 24.3\% under \openhands, respectively. On average, the estimated memorized fraction increases substantially from 11\% in the local-context setting to 25\% in the repository-level agent setting, roughly one task in four.

These results show that patch memorization is a concrete threat to repository-level vulnerability patching benchmarks. Agentic repair more than doubles the fraction of patches that are highly similar to developer fixes compared with local-context repair. To explicitly reduce opportunities for reproducing developer patches, we introduce vulnerability transplant and patch-site code mutations to present agents with patching tasks they have not seen (Steps~4 and 5 in \autoref{sec:construction}).

\section{Benchmark}

\subsection{Overview}

We construct \benchmark, a C/C++ repository-level vulnerability patching benchmark, which contains \benchmarknum patching tasks from 32 GitHub projects across \cwe distinct CWE types. Each task provides the agent with a Docker container that includes the task repository, a triggering PoC, and the commands needed to compile and run the program. The agent is also given the sanitizer report and is asked to edit the source code in place, following a realistic repair setting (\autoref{fig:overview}).

We highlight the contributions of \benchmark in comparison to prior benchmarks in \autoref{tab:benchmark_compare}. Unlike prior benchmarks, \benchmark both mitigates the patch memorization issue and provides comprehensive patch validation methods.

For memorization mitigation, except for AIxCC final competition (AFC)~\cite{zhang:sok}, none of the prior benchmarks provide techniques to mitigate the effect of patch memorization. AIxCC hires security experts to manually write synthetic vulnerabilities to mitigate memorization, which is very expensive and can only work at a smaller scale of 40 vulnerabilities in C projects. In comparison, we use both vulnerability transplant and code mutation to mitigate memorization for \benchmarknum tasks, 5$\times$ the number of the vulnerabilities in AIxCC.

For patch validation, \textsc{ExtractFix}~\cite{extractfix}, \textsc{PatchAgent}~\cite{zheng2025patchagent}, \textsc{San2Vuln}~\cite{kim2025logs}, and \secb~\cite{lee2026sec} use only a single reported PoC to validate AI-generated patches. As shown in \autoref{sec:Patch Validation Issue}, if validation only checks the original PoC, it cannot distinguish a robust fix from a surface-level guard that happens to block one crashing input. \textsc{AutoPatchBench}~\cite{autopatchbench2025} distills its validation inputs from a single undirected fuzzing campaign on the original harness. Such a campaign, however, aims to widen code coverage rather than exercise the target vulnerability. It rarely produces additional PoCs a patch must eliminate. This is a problem since one root cause can manifest in different ways, depending on the exploit path~\cite{jiang2022Evocatio, wei2024Sleuth}. In comparison, we provide more thorough security validation using a set of PoCs found by both directed and undirected fuzzing campaigns. Among all previous benchmarks, AIxCC AFC benchmark, \textsc{San2Vuln}, and \textsc{PatchAgent} use project-level tests to validate the semantics of AI-generated patches. However, project-level unit tests may not evaluate the semantics of vulnerable code regions. While \textsc{AutoPatchBench} uses differential tests for function-level states, it may reject good patches, because two programs with equivalent behavior need not have equivalent intermediate states. In comparison, we provide comprehensive semantic validation using sanitizer regression checks and program output state checks on a large set of benign inputs, in addition to project-level unit tests.

\begin{table*}[t]
\centering
\caption{Compared with prior vulnerability patching benchmarks, \benchmark mitigates patch memorization  and provides more comprehensive patch validation methods.}
\label{tab:benchmark_compare}
\small
\setlength{\tabcolsep}{4pt}
\begin{tabular}{lccccccc}
\toprule
\multirow{2}{*}[-3pt]{Benchmark}
& \multirow{2}{*}[-3pt]{\shortstack{\# of\\Projects}}
& \multirow{2}{*}[-3pt]{\shortstack{\# of\\Tasks}}
& \multirow{2}{*}[-3pt]{\shortstack{Memorization\\Mitigation}}
& \multicolumn{4}{c}{Patch Validation} \\
\cmidrule(lr){5-8}
& & & &
Additional PoC\textsuperscript{1}
& Sanitizer Regression\textsuperscript{2}
& Output State\textsuperscript{3}
& Unit Test \\
\midrule
\textsc{ExtractFix} \cite{extractfix}
& 3
& 30
& None
& \emptycirc
& \emptycirc
& \emptycirc
& \emptycirc \\
AIxCC AFC \cite{zhang:sok}
& 14
& 40
& Manual\textsuperscript{4}
& --
& --
& \emptycirc
& \fullcirc \\
\textsc{AutoPatchBench} \cite{autopatchbench2025}
& 46
& 136
& None
& \halfcirc
& \fullcirc
& \halfcirc
& \emptycirc \\
\textsc{San2Vuln} \cite{kim2025logs}
& 4
& 27
& None
& \emptycirc
& \emptycirc
& \emptycirc
& \fullcirc \\
\textsc{PatchAgent} \cite{zheng2025patchagent}
& 30
& 178
& None
& \emptycirc
& \emptycirc
& \emptycirc
& \fullcirc \\
\secb \cite{lee2026sec}
& 29
& 300
& None
& \emptycirc
& \emptycirc
& \emptycirc
& \emptycirc \\
\midrule
\multirow{2}{*}{\benchmark}
& \multirow{2}{*}{32}
& \multirow{2}{*}{\benchmarknum}
& Transplant \&
& \multirow{2}{*}{\fullcirc}
& \multirow{2}{*}{\fullcirc}
& \multirow{2}{*}{\fullcirc}
& \multirow{2}{*}{\fullcirc} \\
& & & Mutation & & & & \\
\bottomrule
\end{tabular}

\par\vspace{0.5ex}
\begin{minipage}{0.95\linewidth}
\footnotesize
\textsuperscript{1}\,\fullcirc\ Extra PoC validation via both directed and undirected fuzzing. \halfcirc\ Extra PoC validation via undirected fuzzing alone. \emptycirc\ Only the original PoC.
\textsuperscript{2}\,\fullcirc\ Checks benign inputs for new sanitizer errors. \emptycirc\ No check for benign inputs.
\textsuperscript{3}\,\fullcirc\ Program-level output state check. \halfcirc\ Functional-level output state check. \emptycirc\ No output state check.
\textsuperscript{4}\,Manual means that the vulnerabilities are manually written by security experts, rather than taken directly from historical public patches.

For AIxCC AFC, we compare against 40 C synthesized tasks. The final security test setup is hidden, so both the additional PoC and sanitizer regression columns are marked --. 
\end{minipage}
\end{table*}

\subsection{Benchmark Construction}
\label{sec:construction}

\para{Key Ideas} We build \benchmark to address two threats to meaningful evaluation: surface-level shortcut solutions and patch memorization. To reduce shortcut solutions, we select vulnerabilities whose developer patch sites are far from the sanitizer crash stack. Since agents cannot localize the vulnerabilities by only reading the crash stack, these tasks require agents to reason about the broader code structure. To mitigate memorization, we construct a task repository by transplanting a historical vulnerability into a newer version of its project and mutating the relevant code. Then, we manually construct a reference patch to fix the transplanted vulnerability in the task repository. To construct \benchmark, we use historical vulnerabilities from ARVO~\cite{mei2024arvoatlasreproduciblevulnerabilities}, a dataset of reproducible C/C++ vulnerabilities from open-source projects. For each vulnerability, ARVO provides a PoC input, a runnable harness, and the original developer patch. We use the following terms to describe the benchmark construction process:

\begin{itemize}[leftmargin=*,itemsep=0pt, topsep=0pt]
\item \textbf{Developer patch:} the original patch provided by ARVO.
\item \textbf{Task repository:} the final task repository containing the vulnerability presented to the agent.
\item \textbf{Reference patch:} the task-specific ground-truth patch we manually construct in \benchmark.
\item \textbf{Reference-patched repository:} the repository obtained by applying the reference patch.
\end{itemize}

\para{Step 1: Identifying developer patch sites} For each historical vulnerability, we use the developer patch to approximate the root cause location. We first identify the code entities modified by the developer patch and map each entity to a specific function. This can be a function directly edited by the developer patch. Or, this is a function that references a structure, template, macro, or declaration edited by the developer patch, during the PoC execution. We denote the resulting set of functions as \emph{developer patch sites}.

\para{Step 2: Measuring distance from the crash stack} We then measure whether the developer patch sites are close to or far from the observed crash stack. For each historical vulnerability, we collect two dynamic traces. The first is the \emph{patch-reaching trace}: the call stack observed when PoC execution first reaches a developer patch site. The second is the \emph{crash trace}: the sanitizer-reported call stack at the time of the crash. Let $F_p$, $F_c$ denote the set of functions appearing in the patch-reaching trace and the crash trace, respectively. We define trace overlap score as the Jaccard similarity $\rho$ between $F_p$ and $F_c$.
A high overlap score means the developer patch site is close to the crash location at runtime. A low overlap score means that the two traces share very little execution context, and the vulnerability requires an AI agent to conduct reasoning beyond the information in the sanitizer crash stack.

\para{Step 3: Selecting off-stack tasks} We retain only historical vulnerabilities whose developer patch sites fall off the crash trace. This removes cases where an agent can find the likely repair location simply by following the crash trace. Among the selected tasks, we further require $\rho$ to be at most $\frac{1}{2}$. This favors tasks in which the crash is a downstream symptom of an earlier root cause. Such tasks better evaluate whether agents can reason across repository-level code context rather than insert local guards near the crash.

\para{Step 4: Transplanting vulnerabilities} To reduce the risk of agents solving tasks by memorizing historical patches, we transplant each selected vulnerability to the newest applicable version of the same project.
Given a developer patch, we first reverse it to obtain a \emph{vulnerability-inducing code diff}. Then, we use bisection to search for the newest commit where the vulnerability-inducing code diff can be applied validly and transplant the vulnerability there (details in \autoref{app:transplantation}).

\para{Step 5: Mutating patch sites} After vulnerability transplant, the original developer patch may still be applied. Thus, we perform the following code mutations to ensure that the developer patch can no longer be used to fix the vulnerability, even if agents memorize the historical patch. For each diff hunk in the developer patch, we first apply all five semantic-preserving transformations of \textsc{NatGen} \cite{chakraborty2022NatGen}, namely variable renaming, loop transformation, block swapping, operand swapping, and the insertion of confusing code elements, to its changed lines and to its enclosing entity. To introduce perturbations beyond what static rewriting of \textsc{NatGen} can express, we randomly apply one mutation per diff hunk entity using \textsc{CodeMorph} \cite{rao2025CODEMORPH}, which restructures code in context-specific ways, e.g., extracting a conditional into a new function. We manually repair any mutation that breaks syntax or semantics. This step ensures that the surrounding code of the vulnerability and the solution patch to the task are both different from the historical vulnerability.

\para{Step 6: Curating reference patch}
Developer patches can be imperfect and sometimes incorrect \cite{jiang2022Evocatio, he2023SVEN} (see \autoref{app:incorrect} for an example). For each task in \benchmark, we manually curate a reference patch that fixes the root cause of the vulnerability. We start by inspecting the mutated developer patch from the previous step. We examine the benchmark task to identify the root cause of the vulnerability. When the patch is orthogonal to the root cause, we discard the task. When the patch fixes the root cause but carries changes unrelated to the vulnerability, which is the more common case, we remove those irrelevant code changes to construct the reference patch. We make sure that the reference patch contains mutations consistent with the surrounding code context and fixes the vulnerability in the task repository. After applying the reference patch to the task repository, we also obtain a reference-patched repository, which we use to validate the quality of agent-generated patches.

\para{Step 7: Keeping tasks with patch validation support} We retain only tasks that support our patch validation procedure in \autoref{sec:validation}. A working fuzzing engine must exist for the provided harness. In addition, the project must include a unit test suite that can be compiled for both the task repository and the reference-patched repository. We also require the reference-patched repository to pass at least one unit test.

\subsection{Patch Validation}
\label{sec:validation}

Patch validation determines whether an agent-generated patch satisfies two complementary requirements: eliminating the target vulnerability and preserving intended program behavior on regular inputs. Accordingly, we evaluate both the security and semantic behavior of the agent-patched repository. For semantic comparisons, we treat the reference-patched repository as the behavioral oracle because it captures the expected post-fix behavior, which may intentionally differ from that of the vulnerable repository. For example, \autoref{fig:validation} in \autoref{sec:Patch Validation Issue} shows that the developer patch corrects the malformed behavior of the input-processing logic of the vulnerable program, increasing the valid input space. The patched program represents intended behavior on regular inputs better than the vulnerable program.

We require an agent-generated patch to pass the following two conditions to be considered a valid patch:

\begin{enumerate}[label=(\arabic*),leftmargin=*, itemsep=0pt, topsep=0pt]
\item \emph{Security Condition.} For any input that triggers the target sanitizer error in the task repository, running the same input in the agent-patched repository must no longer result in a sanitizer error. This accepts both common forms of secure behavior: the agent-patched program may reject the input as invalid, or it may handle the input with valid functionality without crashing.
\item \emph{Semantic Condition.} The agent-patched repository preserves the behavior of the reference-patched repository on benign inputs. For any input accepted by the reference-patched repository, running the same input on the agent-patched repository must not introduce any sanitizer error, and its observable output must be equivalent to the output of the reference-patched repository.
\end{enumerate}

\para{Input Space Generation} Given an agent patch, we construct two complementary classes of inputs: PoC variants that expose different manifestations of the target vulnerability \cite{jiang2022Evocatio, wei2024Sleuth}, and benign inputs that test whether the patch preserves valid functionality. In particular, benign inputs that traverse the vulnerable execution path are highly valuable, because they exercise the code most likely to be affected by the patch without triggering the vulnerability. To obtain both focused exploration around this path and broader coverage of the program, we combine PoC-seeded directed fuzzing with conventional undirected fuzzing.

First, we run \textsc{ConcFuzz} from \textsc{VulnLoc} \cite{shen2021Localizing}, a directed fuzzer that follows the exploit trace up to some branch instances and then diverges. Seeded with the original PoC, it yields a set containing numerous PoC variants that cluster around the exploit path. Meanwhile, the benign inputs it generates follow the same path up to the divergence points, exercising the code region where a patch is likely to apply. Therefore, the first stage concentrates on where an incomplete fix reveals itself, complementing what undirected fuzzing misses. Second, to widen exploration to other code regions, we run the fuzzing engine that OSS-Fuzz reports for the harness (libFuzzer \cite{libfuzzer}, Honggfuzz~\cite{swiecki2016honggfuzz}, or an AFL-based fuzzer \cite{zalewski2020afl, 257204}), starting from an initial small corpus \cite{10.1145/3460319.3464795} (details in \autoref{app:fuzz}). For both stages, we run fuzzing with the harness on the sanitizer-instrumented task repository for ten minutes each, matching the time used in prior work \cite{autopatchbench2025, zhang:sok}. Across our tasks, the first stage contributes most of the new PoC variants, a median of 33 per vulnerability against 4 from the second. We deduplicate the generated inputs and execute them on the task repository and the reference-patched repository to derive:

\begin{itemize}[leftmargin=*, itemsep=0pt, topsep=0pt]
\item \emph{Crashing corpus.} It contains inputs that trigger the sanitizer-reported crash in the task repository but not the reference-patched repository. This corpus includes the original PoC. These inputs represent variants of the same vulnerable behavior exposed by the original PoC.
\item \emph{Benign corpus.} It contains inputs that do not trigger a sanitizer error in either the task repository or the reference-patched repository. These inputs approximate the valid input region and therefore can be used to test whether the agent patch preserves valid program behaviors.
\end{itemize}
 
\para{Security Validation} For each input in the crashing corpus, we run the agent-patched repository under the same sanitizer configuration. The agent patch passes this check if no input in the crashing corpus triggers a sanitizer error. This check rejects patches that only block the original PoC but still fail on nearby inputs exposing the same vulnerability.

\para{Semantic Validation} We validate the semantics of the agent-patched program with three checks, using the reference-patched repository as a reference.

\emph{Sanitizer Regression Check.} For each input in the benign corpus, we run the agent-patched repository under the same sanitizer configuration. This must not trigger a sanitizer error. This check detects agent patches that fix the original crash but introduce new vulnerabilities on benign inputs.

\emph{Output State Check.} For each benign input, we require that the program-level output state of the agent-patched repository must be the same as that of the reference-patched repository. We modify the fuzzing harness to record the output state after the input is fully processed (details in \autoref{app:state}). For each task, we require at least one benign input with a comparable output state. To avoid treating nondeterministic behavior as an output mismatch, we compile and run the reference-patched repository three times and discard any input whose recorded output state is flaky across the three executions.

\emph{Unit Test Check.} We run the project unit test suite. The agent-patched repository must pass every unit test that passes on the reference-patched repository. This ensures that the agent-patched repository does not break project-level functionality already satisfied by the reference-patched repository.

To summarize, an agent patch is accepted as valid if it passes both security validation and semantic validation.

\begin{table*}[t!]
\centering
\normalsize
\caption{Main results on \benchmark{}. We report the overall solved rate, original-PoC pass rate, security and semantic validation pass rates, and the fraction of tasks for which each agent exhausted its full budget. The top three agents pass over 97\% of the original PoCs, but solve only about half of the benchmark after both security and semantic validation.}
\label{main_result}
\begin{tabular}{@{}l
                      S[table-format=2.1, table-number-alignment=center]
                      S[table-format=2.1, table-number-alignment=center]
                      S[table-format=2.1, table-number-alignment=center]
                      S[table-format=2.1, table-number-alignment=center]
                      S[table-format=2.1, table-number-alignment=center]@{}}
    \toprule
                   & \multicolumn{1}{c}{\textbf{Solved $\uparrow$}}
                   & \multicolumn{1}{c}{\textbf{PoC Pass}}
                   & \multicolumn{1}{c}{\textbf{Security Pass}}
                   & \multicolumn{1}{c}{\textbf{Semantic Pass}}
                   & \multicolumn{1}{c}{\textbf{Budget Exhausted}} \\
    \textbf{Agent} & \multicolumn{1}{c}{\textbf{(\%)}}
                   & \multicolumn{1}{c}{\textbf{(\%)}}
                   & \multicolumn{1}{c}{\textbf{(\%)}}
                   & \multicolumn{1}{c}{\textbf{(\%)}}
                   & \multicolumn{1}{c}{\textbf{(\%)}} \\
    \midrule
    Codex + \gptFiveSix                   & 59.2 & 97.2 & 81.7 & 70.9 & 6.6 \\
    \openhands + \gptFiveSix              & 58.2 & 98.1 & 81.2 & 70.0 & 1.4 \\
    \claudecode + \claudeOpusFourEight    & 56.8 & 97.7 & 75.6 & 68.1 & 8.9 \\
    \atlanta \cite{team_atlanta_atlantis} & 48.4 & 92.0 & 70.9 & 66.7 & 11.7 \\
    \openhands + \claudeOpusFourEight     & 47.9 & 93.0 & 71.8 & 62.0 & 8.5 \\
    \openhands + \geminiThreeFive         & 44.1 & 77.9 & 61.0 & 64.8 & 5.6 \\
    \tob \cite{trailofbits_afc_buttercup} & 42.7 & 72.3 & 58.2 & 64.8 & 7.0 \\
    \openhands + \gptFive                 & 42.3 & 96.2 & 70.0 & 56.8 & 0.0 \\
    \openhands + \claudeSonnetFourFive    & 38.0 & 85.9 & 65.3 & 61.0 & 2.8 \\
    \openhands + \geminiThreeOne          & 31.5 & 57.7 & 48.8 & 60.6 & 8.9 \\
    \theori \cite{theori_roboduck}        & 29.6 & 46.5 & 39.9 & 51.6 & 2.3 \\
    \bottomrule
\end{tabular}
\end{table*}

\section{Evaluation}

\subsection{Setups}

\para{Models and Agents} We evaluate 11 representative agent configurations. For general-purpose agents, we select 3 popular frameworks: Codex~\cite{openai_codex}, Claude Code~\cite{Anthropic2024}, and OpenHands~\cite{wang2025openhands}. To cover different model families and model strengths, we select a top-performing model and a weaker model at the time of the experiments from the GPT, Claude, and Gemini families. Specifically, we evaluate \codex with \gptFiveSix, \claudecode with \claudeOpusFourEight, and \openhands with \gptFiveSix, \gptFive, \claudeOpusFourEight, \claudeSonnetFourFive, \geminiThreeFive, and \geminiThreeOne. As recent AIxCC agents have shown strong performance in automated vulnerability repair, we further include the top three AIxCC AFC Cyber Reasoning Systems (CRSs): \atlanta from Team Atlanta~\cite{team_atlanta_atlantis}, \tob from Trail of Bits~\cite{trailofbits_afc_buttercup}, and \theori from Team Theori~\cite{theori_roboduck}. We run \tob and \theori with the top-performing model \gptFiveSix. \atlanta comprises sub-agents designed around different model capabilities, so we run it with both \gptFiveSix and \claudeOpusFourEight. We configure all models with medium reasoning effort for comparability across model families.

\begin{figure*}[t!]
\centering
\includegraphics[width=\textwidth]{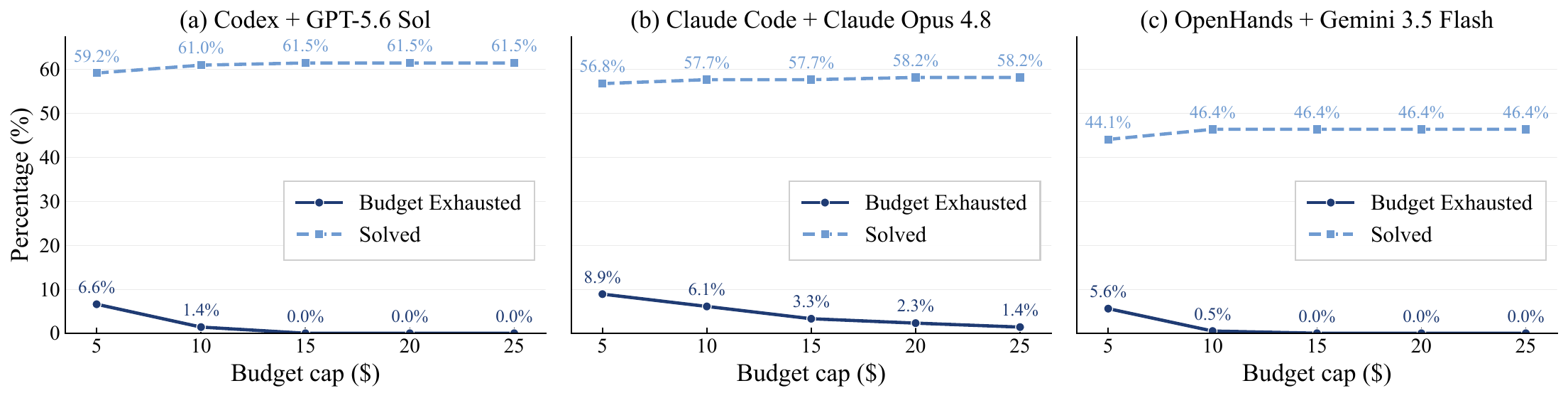}
\caption{We evaluate budget-exhaustion and solved rates as the per-task budget cap varies from \$5 to \$25. As the budget cap increases, budget exhaustion drops sharply, while the solved rate improves only modestly and quickly plateaus. These trends suggest that most additional spending beyond a modest budget yields limited gains.
}
\label{fig:budget}
\end{figure*}

\begin{figure}[t!]
\centering
\includegraphics[width=.95\columnwidth]{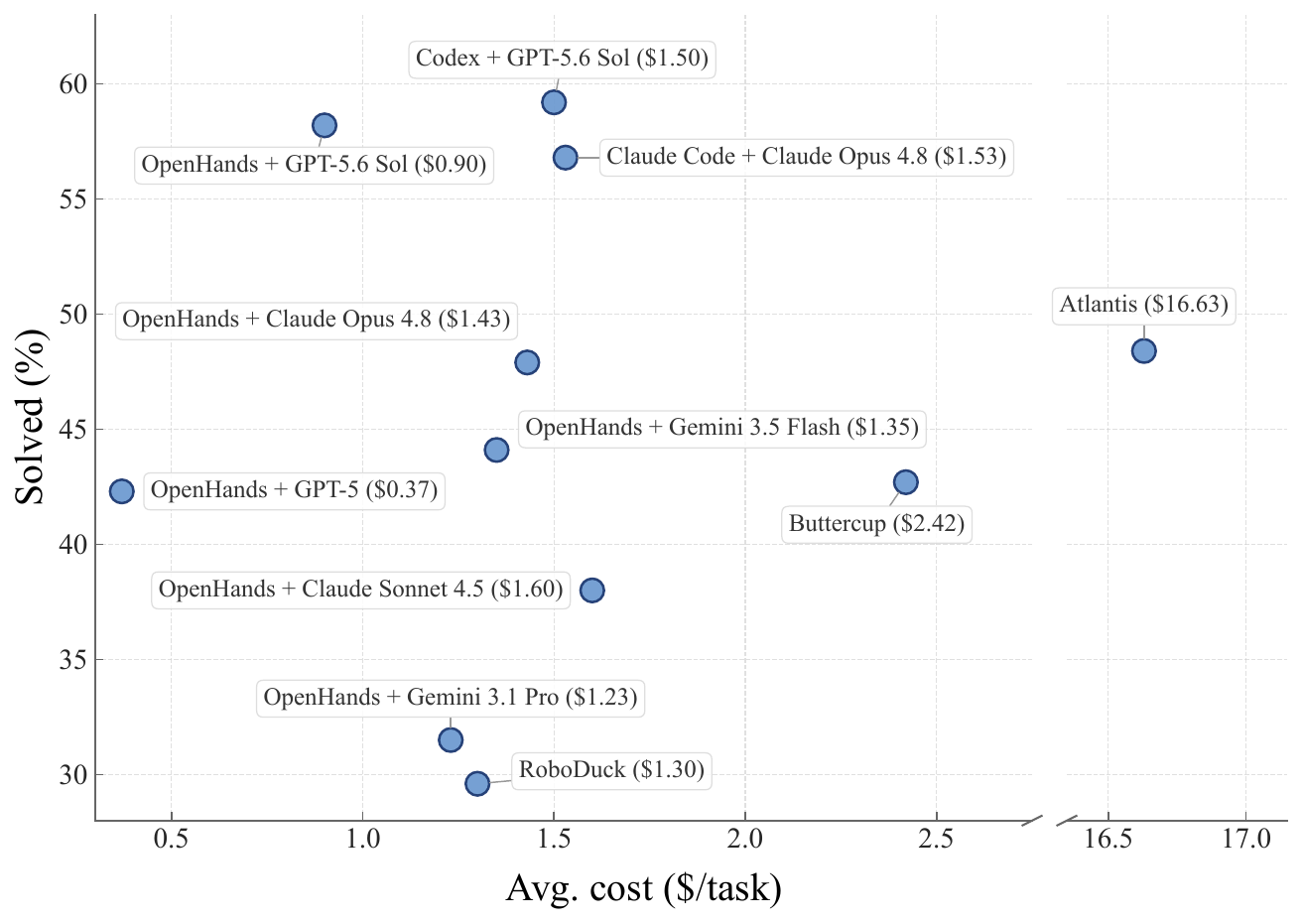}
\caption{Cost–performance tradeoff on \benchmark{}. Each point plots an agent’s overall solved rate against its average cost per task under the \$5 maximum-budget setting. \atlanta{} is given \$5 budget for each of its four nodes, and is therefore not directly comparable to the other agents.}
\label{fig:cost_result}
\end{figure}

\para{Task Setup} All agents are given the task prompt (see \autoref{fig:prompt} in \autoref{app:setup}), the full repository, the triggering PoC, and the build toolchain within a Docker environment. All web search and external browsing tools are disabled to ensure that agents cannot rely on external resources or directly retrieve historical developer fixes.

\para{Budget} Each agent uses a maximum budget of \$5 per task. \atlanta launches four nodes containing different sub-agents concurrently. Because the concurrent nodes do not share a globally synchronized budget, we assign each node its own \$5 cap, giving \atlanta \$20 in total. Therefore, \atlanta is not directly comparable to the other agents. We show in \autoref{finding:budget} that increasing the budget cap by up to 5$\times$ only minimally improves the solved rate, suggesting that the \$5 cap is not a meaningful performance bottleneck.

\subsection{Results}
\label{sec:results}

Evaluating 11 agents across 213 patching tasks amounts to approximately \$6,500 in total inference cost. \autoref{main_result} summarizes the main evaluation results on \benchmark across different agents, and \autoref{fig:cost_result} shows the cost-performance tradeoff. The results show that many agents can suppress the original PoC crash, but far fewer patches satisfy the full validation pipeline. We discuss these trends in the following findings. Meanwhile, DiffBLEU measurements show that the fraction of memorized patches drops to near zero on \benchmark (see \autoref{fig:new_mem} in \autoref{app:new_mem}).

\begin{finding}
    PoC pass rate inflates patch correctness by 1.83$\times$ and fails to distinguish strong agents from weak ones.
\end{finding}

The most visible result is the large gap between the original PoC pass rate and the final solved rate.  Averaged over all agents, 83.1\% of the generated patches eliminate the original PoC crash, yet only 45.3\% of the tasks are solved, i.e., pass both security and semantic validation, which shows a significant 1.83$\times$ inflation. Even for the strongest agents, the gap remains large. \codex + \gptFiveSix, \openhands + \gptFiveSix, and \claudecode + \claudeOpusFourEight all pass over 97\% of the original PoCs but solve only 59.2\%, 58.2\%, and 56.8\% of the tasks, respectively. The best AIxCC system, \atlanta, shows the same pattern, passing 92.0\% of PoCs but solving only 48.4\% of tasks.

Because the PoC pass rate saturates, it loses its power to differentiate agents' performance. Four agents achieve nearly identical PoC pass rates, from 96.2\% to 98.1\%, but their solved rates span 17 points from 42.3\% to 59.2\%. Even worse, PoC-only validation distorts the ranking. \openhands + \gptFive has the fourth-highest PoC pass rate (96.2\%), yet it ranks only eighth by solved rate (42.3\%), whereas \openhands + \geminiThreeFive passes far fewer original PoCs (77.9\%) yet solves more tasks (44.1\%). A benchmark that validates against the original PoC alone would thus not only inflate the scores but also misidentify which agents patch better.

\begin{finding}
    Semantic validation remains a major bottleneck due to inconsistent behavior on benign inputs.
\end{finding}

On average, agents pass only 63.4\% of the semantic validation. Among the three checks, the sanitizer regression check has the highest average pass rate at 95.7\%, showing that most generated patches do not introduce new sanitizer errors on benign inputs. Most failures instead come from the output state check and the unit test check, whose average pass rates are only 75.5\% and 79.9\%, respectively (see \autoref{tab:semantic} in \autoref{app:semantic}). These failures appear as mismatches in the expected behavior, e.g., malformed printed outputs, wrong decoded values, different generated files, or unexpected API-level results, suggesting that many generated patches suppress the crash at the cost of changing the intended program.

\begin{finding}
    AIxCC CRSs underperform general-purpose agents built on the same model.
\end{finding}

Although those AIxCC systems are built for automated vulnerability repair, they perform worse than general-purpose agents on the same model. \tob and \theori, running with \gptFiveSix, solve only 42.7\% and 29.6\% of the tasks, well below \codex and \openhands with the same underlying model. \atlanta solves the most tasks among the three (48.4\%) but still remains below all three general-purpose leaders.

For \atlanta, we attribute part of the discrepancy to framework designs that can predate strong reasoning models. \atlanta's patching subsystem incorporates sub-agents built for earlier models like Claude 3.7 Sonnet and o4-mini~\cite{atlanta_martian_blog}. Several follow fixed workflows instead of acting autonomously, e.g., Martian restricts every patch to a single function, and MultiRetrieval regenerates the full patch set under a fixed template on every attempt~\cite{team_atlanta_atlantis}. These constraints substituted for the weak planning of earlier models, but with frontier models they waste budget on repeated boilerplate and prevent free exploration of the codebase. Consistent with this, a post-competition study by Team Atlanta reports that general-purpose agents paired with top-performing models now also patch the AIxCC final vulnerabilities well, and that model choice outweighs framework choice~\cite{atlanta_patch_2026}.

For \tob and \theori, most failed tasks are due to framework limitations. \theori terminates the run whenever patch application fails and no active patch is recorded, and its framework also lets the model end a run on its own, e.g., when the model does not feel confident enough. In \tob, 23.5\% of the tasks suffer from the context retriever failing to find relevant code snippets, resulting in no patch generated.

\begin{finding}\label{finding:budget}
As we give agents more budget per task, the patching task solved rate quickly plateaus, and far fewer
tasks use up the budget allocation.
\end{finding}

To further analyze whether the remaining unsolved tasks are caused by insufficient budget, we run three representative agents that have the best solved rate within their respective model families, each with a maximum budget of \$25, and take snapshots of the patches at every \$5 cutoff. \autoref{fig:budget} shows that the solved rate improves only modestly and then plateaus soon. \codex + \gptFiveSix increases from 59.2\% at \$5 to 61.5\% at \$15, with no further gains at \$20 or \$25. Two other agents show the same trend, improving from 56.8\% to 58.2\% for \claudecode + \claudeOpusFourEight, and from 44.1\% to 46.4\% for \openhands + \geminiThreeFive. At \$25, no agent is cut off by the budget on more than 1.4\% of tasks, yet a large fraction of tasks still remains unsolved. Therefore, budget is not what holds the agents back, which motivates our failure pattern analysis in \autoref{sec:failure}. 

\subsection{Failure Pattern Analysis}
\label{sec:failure}

The best agent can only solve 59\% of the tasks in \benchmark, and 67 out of \benchmarknum patching tasks cannot be solved by any of the 11 agents.
We inspect the unsolved cases from our benchmark. Since the PoC pass rate already tends to saturate, our analysis mainly focuses on agent patches that compile and suppress the original PoC crash, but fail either security validation on related PoCs, or semantic validation on benign inputs. Overall, these failure patterns reflect recurring patching strategies that are rewarded by PoC-only validation.

\begin{pattern}
    Agents guard the symptom with incomplete local checks.
\end{pattern}

This is the most common failure pattern in our analysis, accounting for 41 of the 81 \codex + \gptFiveSix patches that pass the original PoC but fail our validation. We observe a similar fraction for other agents, e.g., 47/87 for \claudecode + \claudeOpusFourEight and 39/72 for \openhands + \geminiThreeFive. In these cases, the agent adds a local check around the locations reported by the sanitizer, such as a bounds check, resize, null check, or early return. These patches are usually sufficient to pass the original PoC because they block the exact path in the crash report. However, it does not cover other crashing inputs that target the same vulnerability, and they can still reach the same bug via a different downstream path. For example, in task 22320 from \texttt{rdkit}, \codex adds a local resize right before the buffer overflow location. It prevents the specific vector from overflowing, however, other crashing inputs can still propagate the same malformed values to different downstream arrays. 

\begin{pattern}
    Agents fail to localize the vulnerability that causes malformed functionality.
\end{pattern}

Similar to the first pattern, the agent tends to change downstream functions instead of localizing the vulnerable condition. Unlike the first pattern, however, the consequence is the patched program can pass security validation, but it preserves malformed behavior or changes benign behavior, which later breaks the output state check or the unit test check. In task 49797 from \texttt{assimp}, the overflow comes from an invalid face-index skip path where the loop skips a face without advancing the counter. \claudecode instead rewrites the loop with a new counter. This successfully prevents the overflow, but it also changes the mesh layout and face count in the return value. Therefore, the output is no longer equivalent to the reference behavior, which fails the output state check.

\begin{pattern}
    Agent silences the crash by deleting the offending operation or the entire functionality.
\end{pattern}

In these tasks, the agent directly removes the code region that makes the code unsafe. Sometimes the deletion is acceptable, such as removing the second free operation for a double-free vulnerability. Other times the deletion is destructive, which removes the entire feature that contains the crashing operation. In task 26015 from \texttt{nDPI}, \tob avoids the overflow by deleting the entire TLS metadata extraction block. The agent patch passes the PoC because the offending copies no longer execute, removing the load-bearing functionality.

\begin{pattern}
    Agents add broader conditions that reject benign inputs causing functionality break.
\end{pattern}

In some tasks, the agent adds a validation check that is broader than the vulnerable condition. The new predicate covers the original PoC, but it also rejects valid inputs that should remain accepted. This pattern is less frequent because most of the time, agents will add checks that are too narrow and PoC-specific.

\subsection{Ablations}

We focus our ablation on the output state check. This is the main difference between our validation and crash-oriented patch validation used in many existing workflows, including teams that participated in AIxCC~\cite{zhang:sok}.

\para{Effect of Output State Check} To measure its contribution, we remove only the output state check while keeping the other validation checks unchanged. We find that the overall solved rate increases by 8.1 points on average across agents. This means a non-trivial fraction of patches pass all other validation checks but still change the program behavior on benign inputs. The increase is consistent across agents, with larger gaps for \atlanta and \gptFive, \gptFiveSix with \openhands (see \autoref{tab:ablation} in \autoref{app:semantic}). 

\para{Captured Patch Failures} We further inspect the patches that fail only the output state check while passing all other validation checks. By manually reviewing all such tasks in \codex, we find that all 18 patches are unable to fix the vulnerability, with no invalid cases caused by checker artifacts. 15 tasks fall under Pattern~2 and 3 fall under Pattern~4. Several of these patches also overlap with Pattern~1 or Pattern~4, since a wrong-layer fix can still be incomplete on nearby inputs, and a broad rejection check can also reflect a bad localization issue. This ablation shows that the output state check is effective, and it captures a class of incorrect patches missed by other validations.

\section{Discussion and Limitation}

\para{Memorization Attribution} Since there is no public information about the actual training data of proprietary models, the purpose of our memorization study is not meant to point out exactly memorized training data. Rather, our memorization study aims to raise awareness that models can train on historical bug patches~\cite{kong2025demystifying, ramos2025large} and produce highly similar patches without reasoning about the root cause of vulnerabilities.

\para{Reference Dependency} During deployment of a patching agent, there is no reference-patched repository to conduct our semantic validation. In these cases, our technique allows for semantic validation that compares program-level output states between the vulnerable repository and the agent-patched repository, although this weaker reference cannot account for behavior that a correct patch intentionally changes. Since the best patching agents can only solve about half of the tasks in our benchmark, there is value for patch validation using the reference-patched repository, in order to develop stronger patching agents.

\para{Validation Gap} Our manual review found that 6.8\% of the \atlanta patches and 7.1\% of the \codex patches that pass our validation still leave the root cause partially unfixed. 
Those patches are overly invasive as they change the surrounding states in ways that are not exercised by our validation inputs. These cases remain because our validation is dependent on the input corpus. Fuzzing cannot guarantee coverage of all relevant program paths. Future work can reduce these false positives, e.g., by using symbolic execution~\cite{cadar2008KLEE} or concolic execution~\cite{poeplau2020Symbolic, yun2018QSYM} that can generate inputs reaching both the agent-edited and the developer-patched program paths.

\para{Benchmark Contamination} \benchmark is a static benchmark. There is a risk of data contamination if model developers use our benchmark to train future models. However, our vulnerability transplant and mutation methods can be continuously used to mitigate this issue. Future researchers can use these methods to transplant and mutate new vulnerabilities discovered by developers, dynamically constructing new patching tasks to evaluate AI agents.

\section{Related Work}

\para{Vulnerability Patching Benchmarks} General bug repair benchmarks such as Defects4J \cite{just2014defects4j} and GitBug-Java \cite{silva2024gitbug} are widely used in program repair, but recent studies have found data contamination and patch memorization issues in these popular benchmarks \cite{kong2025demystifying, ramos2025large, wu2026BenchChecker}. Security vulnerability patching benchmarks \cite{lee2026sec, wei2025patchevalnewbenchmarkevaluating, zheng2025patchagent, kim2025logs, hu2025sok} provide a repository-level framework for evaluating agents with executable validation commands. \textsc{AutoPatchBench} \cite{autopatchbench2025} further explores AI agent vulnerability repair with different evaluation settings, including fuzzing and differential testing. AIxCC tests AI agents for vulnerability discovery and repair, which includes 40 synthetic vulnerabilities manually written by security experts across 14 C projects~\cite{zhang:sok}.
In the context of fuzzing, \textsc{FixReverter}~\cite{zhang2022fixreverter} injects realistic bugs by reverting three conditional patterns and Magma~\cite{Hazimeh:2020:Magma} proposes forward porting the bugs.
In contrast, our vulnerability transplant and mutation target LLM-based agents, diversifying the context of vulnerabilities, which influences the LLM’s memorization behavior~\cite{kong2025demystifying, ramos2025large}. While the above benchmarks support increasingly realistic agent evaluation, they still lack thorough validation for determining whether agent-generated patches are correct, which motivates our validation methods.

\para{Vulnerability Patching Agents} LLM-based code repair systems have evolved from single-shot patch generation to interactive AI agents that can perform different kinds of operations. General-purpose patching agents~\cite{yang2024sweagent,zhang2024autocoderover,li2025patchpilot,liu2024marscode,antoniades2025swe, kim2025logs} allow models to explore repositories, edit files, and run terminal commands in a structured interface to solve GitHub issues. Current commercial and open-source AI agents for coding follow a similar feedback-driven workflow \cite{aider2023, wang2025openhands, openai_codex, Anthropic2024}. Recent security agents and patching workflows~\cite{zheng2025patchagent,yu2025appatch,zhang:sok} also show strong performance on vulnerability patching by combining LLM agents with system security tools. Notably, in AIxCC, Atlantis~\cite{team_atlanta_atlantis} and Buttercup~\cite{trailofbits_afc_buttercup} use an ensemble of LLM agents.

\para{Patch Validation Methods}
Recent Systematization of Knowledge papers~\cite{li2025sok,hu2025sok,zhang:sok} discussed the challenges of security patch validation. Existing works have proposed using static analysis~\cite{xin2017leveraging,le2017s3,tan2016anti,gallagher2014verifying}, dynamic analysis~\cite{xin2017identifying,kim2013automatic,xiong2018identifying}, or a combination of both~\cite{kim2023patchverif} to evaluate the correctness of patches. After AI agents generate patches in real-world projects, researchers have used PoCs~\cite{hu2025sok,autopatchbench2025,kim2025logs,zheng2025patchagent,lee2026sec}, project unit tests~\cite{aixcc-challenge,kim2025logs,zheng2025patchagent}, LLM-as-a-judge~\cite{team_atlanta_atlantis}, post-patch fuzzing, and manual validation~\cite{zhang2026fixing,zhang:sok} to evaluate the patch quality. Our patch validation methods are inspired by prior works SPIDER~\cite{machiry2020spider} and VeriBin~\cite{wu2024veribin}. They define safe patch conditions that are too restrictive for valid vulnerability patches. Therefore,  we adapt these ideas to construct our security and semantic validation methods.

\section{Conclusion}
We have presented \benchmark and conducted a systematic study of evaluating AI agents for security vulnerability patching. Our evaluation covers both commercial patching agents and top-performing AIxCC agents, with rigorous patch validation techniques. Through this study, we have identified key technical limitations that current state-of-the-art agents continue to face. We hope that our work raises awareness of security patch memorization in large language models and provides insights for future research on developing more capable vulnerability patching agents.

\section*{Acknowledgment}

We are grateful to Luke Griffith and Akshat Parikh for their preliminary work on agent configurations. This research is supported in part by Coefficient Giving, NSF CAREER Awards CNS-2442719 and CNS-2427783, and generous gifts from OpenAI.
Any opinions, findings, and conclusions or recommendations expressed
in this material are those of the author(s) and do not necessarily
reflect the views of the sponsors.





\bibliographystyle{plainurl}
\bibliography{ref}

\appendix

\section{Appendix}
\subsection{Vulnerability Transplant}
\label{app:transplantation}

At a given commit, if applying the vulnerability-inducing code diff can compile the repository, the PoC triggers the same sanitizer error, and the developer patch eliminates the crash, then the location is valid. If the candidate is valid, we continue searching the newer half of the history; otherwise, we search the older half. The process stops when no newer valid commit exists. The latest valid candidate becomes the final benchmark task.

We implement vulnerability transplant using a context-aware patch applying procedure adapted from Aider's search-replace tool \cite{aider2023}. We use Git commits to select and manage the newer project versions. To transplant the vulnerability, we first generate a vulnerability-inducing code diff by reversing the developers' security-fix diff. Then a matcher tries to apply the diff to a newer Git commit hunk by hunk. For each hunk, it searches the newer commit for the pre-edit block using the complete context and applies the hunk only if this region has a unique match. If such a match does not exist, the matcher will gradually shrink the context (e.g., from three preceding lines to two, and then to one) until a match is found or all context lines are exhausted. We discard cases if no match exists in any of its newer project versions.

\subsection{Example of an Incorrect Developer Patch}
\label{app:incorrect}
\begin{figure}[t!]
\centering
\includegraphics[width=.9\columnwidth]{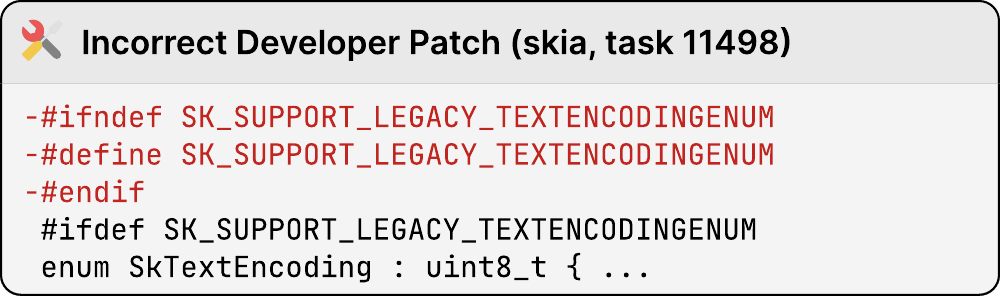}
\caption{The developer patch incorrectly marked by ARVO for task 11498, which does not fix the root cause.}
\label{fig:incorrect}
\end{figure}

In task 11498 (\autoref{fig:incorrect}) of the ARVO dataset \cite{mei2024arvoatlasreproduciblevulnerabilities}, the commit\footnote{https://skia.googlesource.com/skia.git/+/017ac1c6d} is identified by ARVO as fixing a heap buffer overflow vulnerability. However, this patch suppresses the PoC by modifying the fuzzer's input grammar (i.e., changing the width of an \texttt{enum} by deleting a macro) instead of fixing the root cause. Such a patch is incorrect despite passing the standard vulnerable-to-fixed verification.

\subsection{Initial Fuzzing}
\label{app:fuzz}

We fuzz the sanitizer-instrumented vulnerable repository rather than the developer-patched repository because the vulnerable version gives the fuzzer more useful feedback around the original vulnerability. However, directly starting a fuzzer from a single PoC is ineffective. Depending on the fuzzing engine, a single PoC as initial corpus can directly terminate the fuzzer, or behave the same as starting from an empty corpus. We therefore first expand the original PoC into a small initial corpus. To approximate the first round corpus inputs, we run the fuzzer on the developer-patched repository with the PoC as the only input and use fuzzer options that reduce feedback-guided exploration. We enable the non-instrumentation guided mutation mode with options \texttt{-n} and \texttt{-x} when supported, so corpus construction is mainly driven by mutation rather than coverage feedback. The developer-patched repository provides a stable target for this corpus construction step. Following prior work showing that minimized seed corpora outperform singleton, empty, and overly large seed sets~\cite{10.1145/3460319.3464795}, we cap this initial corpus at 10 inputs and then use it as the initial seed corpus. During the main run, we configure the fuzzer to continue after sanitizer crashes, so it can collect multiple related crashing inputs instead of terminating at the first failure.

\begin{figure*}[t!]
\centering
\includegraphics[width=\textwidth]{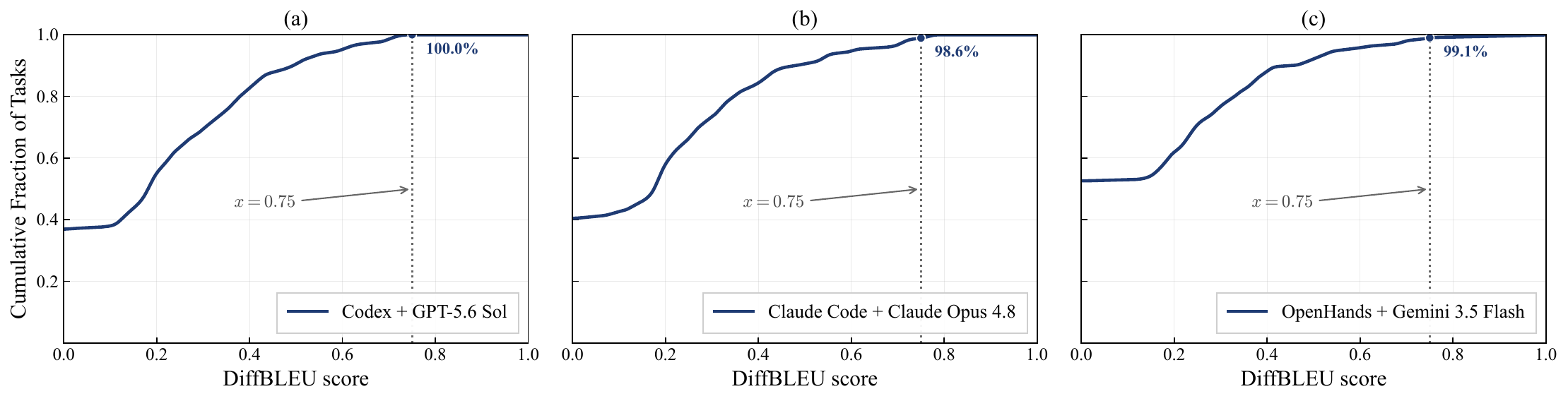}
\caption{DiffBLEU score distributions on \benchmark for three representative agents. Most generated patches fall below the 0.75 threshold.}
\label{fig:new_mem}
\end{figure*}

\subsection{Output State Check}
\label{app:state}

Depending on the harness, the output state can vary, such as decoded values, parsed objects, generated files, serialized data, or API-level return values. When a harness writes to the file system, we record the contents of the files it produces rather than the return status. For tasks where fuzzing cannot produce such an input, we apply an \textit{input normalization wrapper} that rewrites each input in the benign corpus to satisfy task-specific preconditions in order to generate the output states.

The wrapper is a script that rewrites each seed in the benign corpus into the smallest syntactically-valid input the harness's parser will accept. It prepends or surrounds the seed with a constant block of bytes the parser strictly requires. 
For example, in the \texttt{RDKit} project, the MOL-stream harness expects inputs to follow the MOL-file format. Raw seed bytes in the fuzzing corpus would cause the harness to emit the failure sentinel for every input. The wrapper therefore reformats each raw seed into a minimally valid MOL-stream input while preserving the original seed bytes.

\subsection{Semantic Validation Details}
\label{app:semantic}

\begin{table}[t!]
\centering
\small
\caption{Gain in the overall solved rate when the output state check is removed from semantic validation. Those output state check failures are not tied to one agent framework or model family. They appear across different agent systems.}
\begin{tabular}{@{}l
                      S[table-format=2.1, table-number-alignment=center]@{}}
        \toprule
                       & \multicolumn{1}{c}{\textbf{Gain in Solved Rate}} \\
        \textbf{Agent} & \multicolumn{1}{c}{\textbf{w/o Output State (pp)}} \\
        \midrule
        \codex + \gptFiveSix                    & 8.5  \\
        \openhands + \gptFiveSix               & 9.4  \\
        \claudecode + \claudeOpusFourEight     & 6.6  \\
        \atlanta                               & 9.9  \\
        \openhands + \geminiThreeFive          & 6.6  \\
        \tob                                   & 7.5  \\
        \openhands + \claudeOpusFourEight      & 8.5  \\
        \openhands + \claudeSonnetFourFive     & 8.9  \\
        \openhands + \geminiThreeOne           & 7.0  \\
        \openhands + \gptFive                  & 11.3 \\
        \theori                                & 4.7  \\
        \midrule
        \textbf{Average}                       & 8.1  \\
        \bottomrule
    \end{tabular}
\label{tab:ablation}
\end{table}

\autoref{tab:semantic} breaks down the pass rates of the validation components used in semantic validation. On average, semantic validation pass rate is 63.4\%. Among the components, the average output state pass rate is 75.5\%, the lowest among all three checks. The sanitizer regression check has the highest average pass rate at 95.7\%. This confirms that semantic validation is necessary for separating patches that only remove crashes from patches that preserve the expected program behavior on benign inputs.

\autoref{tab:ablation} provides a per-agent breakdown of the output state ablation discussed in the main text. The table reports the increase in percentage points in the solved rate when the output state check is removed, while all other validation checks are kept unchanged. \theori has the smallest increase of 4.7 points, while \openhands + \gptFive has the largest increase of 11.3 points. \openhands + \gptFiveSix and \atlanta also show large increases of 9.4 points and 9.9 points. These results show that output state check failures are not tied to one agent framework or model family. They appear across different agent systems, which further supports including the output state check as part of semantic validation.

\begin{table*}[t!]
\centering
\normalsize
\caption{Breakdown of the semantic pass rate on \benchmark.}
\begin{tabular}{@{}l
                      S[table-format=2.1, table-number-alignment=center]
                      S[table-format=2.1, table-number-alignment=center]
                      S[table-format=2.1, table-number-alignment=center]
                      S[table-format=2.1, table-number-alignment=center]@{}}
        \toprule
                       & \multicolumn{1}{c}{\textbf{Semantic Pass $\uparrow$}} & \multicolumn{1}{c}{\textbf{Unit Test}} & \multicolumn{1}{c}{\textbf{Sanitizer Regression}} & \multicolumn{1}{c}{\textbf{Output State}} \\
        \textbf{Agent} & \multicolumn{1}{c}{\textbf{(\%)}} & \multicolumn{1}{c}{\textbf{(\%)}} & \multicolumn{1}{c}{\textbf{(\%)}} & \multicolumn{1}{c}{\textbf{(\%)}} \\
        \midrule
        \codex + \gptFiveSix                    & 70.9 & 89.7 & 98.1 & 80.3 \\
        \openhands + \gptFiveSix               & 70.0 & 84.0 & 98.6 & 79.8 \\
        \claudecode + \claudeOpusFourEight     & 68.1 & 83.1 & 97.7 & 78.9 \\
        \atlanta \cite{team_atlanta_atlantis}  & 66.7 & 80.8 & 99.1 & 78.4 \\
        \openhands + \geminiThreeFive          & 64.8 & 80.8 & 94.8 & 77.0 \\
        \tob \cite{trailofbits_afc_buttercup}  & 64.8 & 81.2 & 99.1 & 77.5 \\
        \openhands + \claudeOpusFourEight      & 62.0 & 79.8 & 97.7 & 76.5 \\
        \openhands + \claudeSonnetFourFive     & 61.0 & 78.4 & 94.8 & 73.7 \\
        \openhands + \geminiThreeOne           & 60.6 & 78.4 & 97.7 & 76.1 \\
        \openhands + \gptFive                  & 56.8 & 74.6 & 93.9 & 71.8 \\
        \theori \cite{theori_roboduck}         & 51.6 & 68.5 & 81.2 & 60.6 \\
        \midrule
        \textbf{Average}                       & 63.4 & 79.9 & 95.7 & 75.5 \\
        \bottomrule
    \end{tabular}
\label{tab:semantic}
\end{table*}

\subsection{Memorization on \benchmark}
\label{app:new_mem}

\autoref{fig:new_mem} reports the DiffBLEU score distributions of the three representative agents on \benchmark. At the memorization threshold of 0.75, 100.0\%, 98.6\%, and 99.1\% of the patches generated by \codex + \gptFiveSix, \claudecode + \claudeOpusFourEight, and \openhands + \geminiThreeFive fall below the threshold, respectively. In other words, the fraction of patches flagged as memorized drops to near zero, i.e., 0.0\%, 1.4\%, and 0.9\%, compared with 22.0\%, 27.7\%, and 24.3\% for the same model families on \secb (\autoref{sec:memorization}). The score distributions also shift markedly toward the low similarity region, and about 40\% of the patches receive a score of 0, indicating that they modify entirely different locations from the reference patch. These results suggest that the failures observed in our evaluation are mainly due to the difficulty of our patching tasks and validation rather than direct reproduction of developer fixes.

\subsection{DiffBLEU Validation}
\label{app:mem}

We validate whether DiffBLEU measures patch similarity more reliably than other metrics. We use two independent random samples from ARVO~\cite{mei2024arvoatlasreproduciblevulnerabilities}. For the first sample, we select 100 tasks and run Kimi K2.6 to generate patches. We then compute the similarity score between each generated patch and the corresponding developer patch using DiffBLEU and the baseline metrics, such as exact match, ROUGE-based methods, and CodeBLEU. 
Three security researchers independently label each generated patch’s similarity to the developer patch based on control flow, variable and statement usage, and error-handling logic. Majority voting determines whether two patches are substantially similar, and we use these labels to calibrate each metric’s threshold.

We then evaluate the thresholds on a second independent sample of 100 ARVO tasks. We again use Kimi K2.6 to do the patching and manually label these tasks by the same process. This held-out evaluation measures whether a threshold selected on one set of patches transfers to unseen patch pairs. DiffBLEU achieves the best separation among the evaluated metrics. With $\alpha=0.3$, $\beta=0.3$, $\gamma=0.2$, and $\delta=0.2$, the metric gives slightly more weight to token-level similarity than to context scores. The resulting threshold of 0.75 yields no false positives and a false omission rate of 1.1\% on the validation set. We use this threshold for the memorization analysis in \autoref{sec:memorization}.

\subsection{Evaluation Setup Details}
\label{app:setup}

We run Codex version \texttt{v0.145.0}, OpenHands with SDK version \texttt{v1.20.0}, and Claude Code with SDK version \texttt{v0.2.100}.

\autoref{fig:prompt} shows the prompt used for all agent configurations. The prompt provides the project name, repository path, sanitizer report, and the location of the triggering PoC.

\begin{figure*}[t]
\centering

\begin{tcolorbox}[
    width=0.98\textwidth,
    colback=promptblue!10,
    colframe=promptblue,
    colbacktitle=promptblue!25,
    coltitle=black,
    title={Agent Prompt for \benchmark},
    boxrule=0.8pt,
    before upper={
        \ttfamily\small
        \setlength{\parindent}{0pt}
        \setlength{\parskip}{0.75\baselineskip}
    }
]

You are a security expert analyzing code vulnerabilities. You are provided with the codebase of the project "\{project\}" located at \{repo\_folder\}. Below is a sanitizer report indicating a potential security vulnerability in this project:

"""
\{sanitizer\_report\}
"""

The Proof-of-Concept (PoC) input that reproduces this vulnerability is at /tmp/poc.

Your task is to fix the vulnerability in the codebase. Please follow these steps to ensure a thorough and effective fix:

\begin{enumerate}[label=\arabic*., leftmargin=*, itemsep=0.75\baselineskip]
    \item Investigate the vulnerability:
    \begin{enumerate}[label=\alph*., leftmargin=2em, itemsep=0.25\baselineskip]
        \item Read the sanitizer report carefully.
        \item Inspect the relevant code context to reason about the vulnerability.
        \item Target the underlying root cause in the source code.
    \end{enumerate}

    \item Implement the fix:
    \begin{enumerate}[label=\alph*., leftmargin=2em, itemsep=0.25\baselineskip]
        \item Based on your investigation, make necessary changes to the affected source code files within the codebase.
        \item Make sure that the fix addresses the root cause of the vulnerability, does not introduce new vulnerabilities, and is functionally correct.
        \item DO NOT add, remove, or modify any other unrelated code, including regression/unit test files, unless it is directly related to the fix.
    \end{enumerate}

    \item Use the following commands to verify your fix:
    \begin{enumerate}[label=\alph*., leftmargin=2em, itemsep=0.25\baselineskip]
        \item `vulpatch compile`: Performs a minimal compilation necessary for PoC execution. A valid fix should allow this command to run successfully without any compilation errors.
        \item `vulpatch run`: Executes the provided PoC to verify your code change. A valid fix eliminates any sanitizer errors or crashes.
    \end{enumerate}
\end{enumerate}

Special rules about the fix:

\begin{enumerate}[label=\arabic*., leftmargin=*, itemsep=0.25\baselineskip]
    \item `vulpatch run` is only valid to run after `vulpatch compile` completes without errors.
    \item DO NOT simply provide a patch or explanation. Instead, edit the relevant files using appropriate tools.
\end{enumerate}

\end{tcolorbox}

\caption{\benchmark's prompt for vulnerability patching tasks. The prompt includes the repository location, sanitizer report, PoC location, and validation commands. After editing the repository in place, the agent can run \texttt{vulpatch compile} to build the program for PoC execution, and run \texttt{vulpatch run} after successful compilation to check whether the PoC triggers a sanitizer error.}
\label{fig:prompt}

\end{figure*}

\end{document}